\RequirePackage{rotating}
\documentclass[usenatbib]{mnras}
\usepackage[utf8]{inputenc}
\usepackage{multirow}
\usepackage{rotating}
\usepackage{graphicx}
\usepackage{booktabs}
\usepackage{textcomp}
\usepackage{pbox}
\usepackage{natbib}
\usepackage{mathtools}
\usepackage{gensymb}
\usepackage{colortbl}
\usepackage{color}
\usepackage{amssymb}
\usepackage{amsfonts}
\usepackage{verbatim}
\usepackage{scalefnt}
\usepackage[percent]{overpic}
\usepackage[T1]{fontenc} 
\usepackage{aecompl} 
\usepackage{ulem}
\usepackage{cleveref}
\usepackage{todonotes}
\usepackage[english]{babel}
\usepackage{siunitx}
\usepackage{pbox}
\usepackage{float}
\usepackage{dblfloatfix}
\usepackage{lineno}
\usepackage{enumitem}
\usepackage{tikz}
\usepackage{tikz-3dplot}
\usetikzlibrary{angles, quotes}
\usepackage{tkz-euclide}
\usepackage{subcaption}
\usepackage{multicol}
\usepackage[]{threeparttable}
\usepackage{siunitx}
\usepackage{txfonts}
\usepackage{lmodern}
\def\lsim{\mathrel{\rlap{\lower 3pt \hbox{$\sim$}} \raise 2.0pt \hbox{$<$}}}
\def\gsim{\mathrel{\rlap{\lower 3pt \hbox{$\sim$}} \raise 2.0pt \hbox{$>$}}}

\graphicspath{{./}{figures/}}

\title{Empirical Structure Models of Uranus and Neptune}

\author[Neuenschwander \& Helled]{Benno A. Neuenschwander$^{1}$\thanks{Contact e-mail: \href{mailto:benno.neuenschwander@outlook.com}{benno.neuenschwander@outlook.com}}, Ravit Helled$^{1}$
\\
$^{1}$ Center for Theoretical Astrophysics and Cosmology, Institute for Computational Science, University of Zurich, \\Winterthurerstrasse 190, CH-8057 Z{\"u}rich, Switzerland}

\date{Received 3 January 2022; Revised 23 February 2022; Accepted 3 March 2022}

\begin{document}
\maketitle

\begin{abstract}
Uranus and Neptune are still poorly understood. Their gravitational fields, rotation periods, atmosphere dynamics, and internal structures are not well determined. 
In this paper we present empirical structure models of Uranus and Neptune where the density profiles are represented by polytropes. By using these models, that are set to fit the planetary gravity field, we predict the higher order gravitational coefficients $J_6$ and $J_8$ for various assumed rotation periods, wind depths, and uncertainty of the low-order harmonics. We show that faster rotation and/or deep winds favour centrally concentrated density distributions. We demonstrate that an accurate determination of $J_6$ or $J_8$ with a relative uncertainty no larger than $10\%$ could constrain wind depths of Uranus and Neptune. We also confirm that the Voyager rotation periods are inconsistent with the measured shapes of Uranus and Neptune.   
We next demonstrate that more accurate determination of  the gravity field can significantly reduce the possible range of internal structures.
Finally, we suggest that an accurate measurement of the moment of inertia of Uranus and Neptune with a relative uncertainty of $\sim1\%$ and $\sim0.1\%$, could constrain their rotation periods and depths of the winds, respectively. 
\end{abstract}


\section{Introduction}    

Uranus and Neptune are the outermost planets of our Solar System. 
Although they represent a unique class of planets, relatively little is known about them \cite[e.g.,][]{Helled2020b, HF2020}. 
Constraining their internal structures and bulk compositions is critical for understanding their formation and evolution  \cite[e.g.,][]{2014prpl.conf..643H,Vazan2018,Scheibe2019,Muller2019,Mousis2020,Bailey_2021,Scheibe2021}.
Structure models are designed to fit the planet's measured mass, radius and gravitation field. The gravitational fields of  Uranus and Neptune have been measured in 1986 and 1989, respectively, when the Voyager II mission flew by these planets \citep{Voy_Uranus,Voy_Neptune}.  
Currently, Voyager II is the only spacecraft that has visited Uranus and Neptune. Accordingly, the data collected incorporate relatively high uncertainties that allow for a relatively broad variety of internal structure models. \\ 
In fact, the name "Ice Giants" might give the impression that the compositions of Uranus and Neptune  consist of ices such as water, methane and ammonia, and have small fractions of rocks. However, the compositions of the planets are poorly constrained. It has been suggested that both planets  could actually be rock-dominated   \cite[e.g.,][and references therein]{Teanby2020,Helled2020b, HF2020}. \\ 
A widely-used method to generate structure models is to first assume the planetary composition and then apply the corresponding physical equations of state. Although planet formation models as well as knowledge of the composition of the solar nebula could be used to guide structure models, the compositions of Uranus and Neptune remain  unconstrained.  
Moreover, modelers often describe the planetary interior assuming  different distinct adiabatic layers, each consisting of constant and well-mixed composition \citep[e.g.,][]{NETTELMANN2013}.
Such layers, however, may not exist in Uranus and Neptune. 
First, formation models suggest that the primordial interiors of giant planets consist of composition gradients \citep{Helled_2017,Valletta2020}. 
Second, rock and water, as well as water and hydrogen can be well mixed inside the planet \citep[e.g.,][]{Soubiran2015,Soubiran2017,Vazan2020b}. Therefore, composition changes are expected to be rather described by composition gradients and modeling the interiors of Uranus and Neptune assuming a three-layer structure might be inappropriate. \\ 
To avoid the potential biases in the assumed planetary composition, one can use empirical structure models instead \citep[e.g.,][]{Marley1995,Podolak2000,Helled2011}. In that case the density profiles are represented by polynomials \cite[e.g.,][]{Helled2011} or polytropes \citep[e.g.,][]{Horedt1983,Neuenschwander_2021}. Although these models are somewhat harder to interpret and lack information about the temperature distribution, they provide an unbiased and potentially a more general approach to describe the planetary structure. Such models therefore also include solutions that correspond to internal structures with composition gradients and/or non-adiabatic temperature profiles.

Regardless of the method used to generate the structure models, in order to model the planetary interior we must rely on (ideally accurate) measurements of the planetary mass, radius, rotation period and gravity field. 
An accurate determination of the planetary rotation period is crucial for structure models because it affects the planet's shape, density profile (and therefore inferred composition) as well as the predicted moment of inertia value \citep[e.g.,][]{Helled2010_prot,NETTELMANN2013}.
Therefore, modeling the interiors of Uranus and Neptune is even more complex given the uncertainty in the rotation periods of the planets:  
measurements of periodicities in  Uranus' and Neptune's radio signals and of their magnetic fields, inferred by Voyager 2, suggest rotation periods of 17.24 h for Uranus \citep{Desch1986} and 16.11 h for Neptune \citep{Warwick1989}, see Table \ref{tab:planet_properties}. We refer to these rotation periods with $P_{\text{Voy}}$. However, it was suggested by  \cite{Helled2010_prot} that these periods do not represent the rotation period of the deep interiors. Searching for the periods that minimize the dynamical heights and are most consistent with the measured planetary shapes,  they suggested modified rotation periods of 16.57 h and 17.46 h for Uranus and Neptune, respectively. 
We refer to these rotation periods with $P_{\text{HAS}}$. \\
In addition, the measured gravity field is also affected by dynamical processes such as winds:
Dynamical processes do not only give rise to non-zero odd $J$ values $J_{n+1}$, but, if strong enough, also affect the even $J$ values $J_{2n}$\footnote{Note that shallow winds only affect higher-order gravitational coefficients ($\geq J_4$), as these are more sensitive to the planetary surface.}.
Observations show for both Uranus and Neptune zonal west winds with $\sim240 $ m$\cdot $s$^{-1}$ and $\sim230$ m$\cdot $s$^{-1}$, respectively, on both hemispheres at latitudes around $\pm 60\degree$ and $\pm 70\degree$, respectively, and strong east winds at Neptune's equator ($\sim380 $m$\cdot$ s$^{-1}$)   \citep{Sromovsky2005,Sromovsky1993}. Note that these wind speeds refer to the Voyager rotation periods and depend on the assumed underlying uniform rotation period. \\

Structure models typically account only for the static part $J_{4}^{stat}$. 
However, the measured gravity field also consists of the dynamical contribution. Therefore, in order to properly model the internal structure and compare the inferred gravity field to the measured one, an estimate of the dynamical contribution is required. 
The effect of the winds on the gravity field of Uranus and Neptune has been investigated in detail in  \cite{Kaspi.2013}.
They split the measured second gravitational harmonic $J_{4}^{meas}$ of Uranus and Neptune into a static part and a dynamical part: $J_{4}^{meas} = J_{4}^{stat} + J_{4}^{dyn}$, where  $J_{4}^{stat}$ and $J_{4}^{dyn}$ are the hydrostatic (static interior with uniform rotation) and dynamic (wind) contributions, respectively. 
\cite{Kaspi.2013} provide the expected ranges for the dynamical contribution $J_{4}^{dyn}$ for both Uranus and Neptune. By comparing the measured gravity fields of the planets with the ones predicted from structure models, it was concluded that winds can penetrate up to 1,100 km in the interior of Uranus and Neptune, which corresponds to $J_{4}^{dyn} = 3 \times 10^{-6}$ for Uranus and $J_{4}^{dyn} = 4 \times 10^{-6}$ for Neptune, respectively. These depths are also consistent with the ones inferred using the ohmic dissipation constraint \citep{Deniz2020}.\\
It should be noted, however, that even with a perfect knowledge of the planet's, size, mass, rotation period, and gravity and magnetic fields, structure models give non-unique solutions. This is due to the degenerate nature of this problem. Here we refer to the entity of possible solutions as the  "solution space". \\

In this paper we investigate the effect of different rotation periods and depths of the winds of Uranus and Neptune on the prediction of $J_6$ and $J_8$, the density profiles, the normalized moment of inertia (MoI) value, and the planetary shape. We then explore how more accurate determinations of $J$ values, MoI or polar radii could be used to further constrain the planet's rotation periods and atmosphere dynamics. 
We also demonstrate that improved measurements of Neptune's $J_2$ and $J_4$, with uncertainties comparable to the ones available for Uranus,   could further  constrain Neptune's predicted $J_6$, $J_8$ and MoI value. 
Our paper is structured as follows: In Section \ref{sec:method} we describe our model assumptions and the calculation method, in Section \ref{sec:results} we present our finding which we summarize and further discuss in Section \ref{sec:summary_and_conclusion}.

\begin{table*}
\begin{threeparttable} 
\centering

 \begin{tabular}{l| c c c l l}
 \hline
\multicolumn{1}{c}{planetary model} & \multicolumn{1}{c}{mass [$M_\oplus$]} & \multicolumn{1}{c}{equatorial radius [km]} & \multicolumn{1}{c}{rotation period [h]} & \multicolumn{1}{c}{$J_2$ [$\times 10^6$]} & \multicolumn{1}{c}{$-J_4$ [$\times 10^6$]} \\
 \hline
Uranus              & 14.536\tnote{a} & 25559\tnote{c} & 17.24\tnote{f} & $3510.68 \pm 0.7$\tnote{a} & $34.17 \pm 1.3$\tnote{a} \\
Uranus--            & 14.536\tnote{a} & 25559\tnote{c} & 16.57\tnote{e} & $3510.68 \pm 0.7$\tnote{a} & $34.17 \pm 1.3$\tnote{a} \\
Uranus corr dyn     & 14.536\tnote{a} & 25559\tnote{c} & 17.24\tnote{f} & $3510.68 \pm 0.7$\tnote{a} & $37.17 \pm 1.3$\tnote{a} \\
Uranus-- corr dyn   & 14.536\tnote{a} & 25559\tnote{c} & 16.57\tnote{e} & $3510.68 \pm 0.7$\tnote{a} & $37.17 \pm 1.3$\tnote{a} \\
Neptune             & 17.148\tnote{b} & 24766\tnote{d} & 16.11\tnote{g} & $3535.94 \pm 4.5$\tnote{b} & $35.95 \pm 2.9$\tnote{b} \\
Neptune+            & 17.148\tnote{b} & 24787\tnote{e} & 17.46\tnote{e} & $3529.95 \pm 4.5$\tnote{b} & $35.82 \pm 2.9$\tnote{b} \\
Neptune corr dyn    & 17.148\tnote{b} & 24766\tnote{d} & 16.11\tnote{g} & $3535.94 \pm 4.5$\tnote{b} & $39.95 \pm 2.9$\tnote{b} \\
Neptune+ corr dyn   & 17.148\tnote{b} & 24787\tnote{e} & 17.46\tnote{e} & $3529.95 \pm 4.5$\tnote{b} & $39.82 \pm 2.9$\tnote{b} \\
Neptune $J_{\text{pre}}$   & 17.148\tnote{b} & 24766\tnote{d} & 16.11\tnote{f} & $3535.94 \pm 0.7$\tnote{b}          & $35.95 \pm 1.3$\tnote{b} \\
\hline
\end{tabular}
\caption{Planetary properties of various \textit{planetary models}. 
For Neptune, the equatorial radius and $J$ values are adjusted for the corresponding rotation period \citep[see e.g.,][]{Helled2010_prot}.
$^a$\protect\cite{Jacobson.2014}, $^b$\protect\cite{Jacobson.2009}, $^c$\protect\cite{Lindal1987}, $^d$\protect\cite{Lindal1992}, $^e$\protect\cite{Helled2010_prot}, $^f$\protect\cite{Desch1986},
$^g$\protect\cite{Warwick1989}
}
\label{tab:planet_properties}
\end{threeparttable}
\end{table*}

\section{Methods} \label{sec:method}

The internal planetary structure can not be observed directly, but must be inferred via theoretical models that are designed to fit the planet's observable data such as its total mass $M$, equatorial radius $a$, rotation period and gravitational field. 
Assuming that the planet is in hydrostatic equilibrium (HE) its equilibrium shape can be inferred by using the planetary total potential $U(\Vec{r})$ that consists of the the gravitational $V(\Vec{r})$ and centrifugal  $Q(\Vec{r})$ potentials. The total potential is given by:
\begin{align}\label{eq:total_potential}
    U(r, \theta) &= V(r, \theta) + Q(r, \theta) \nonumber \\
    &= -\frac{GM}{r} \left(1-\sum_{n=1}^{\infty}\left( \frac{a}{r}\right)^n J_n P_n(\cos\theta)\right) \\
    &\quad\text{ }+ \frac{1}{2}\omega^2r^2\sin^2(\theta) \nonumber,
\end{align}
where $r$ and $\theta$ are the distance and co-latitude, respectively,  
$G$ is the gravitational constant, $J_{2n}$ are the gravitational coefficients (or $J$ values), $P_n(\cos\theta)$ the Legendre polynomials and $\omega$ the rotation rate. \\
The gravitational coefficients $J_n$ can be calculated as an integral over the radial density profile $\rho(r)$, given the planetary volume $\tau$:
\begin{equation}
    Ma^nJ_n = - \int_\tau \rho(r)r^nP_n(\cos\theta)d\tau
\end{equation}
$\tau$ is the volume enclosed by the planetary surface that itself is described by the equipotential surface (i.e. $U(r, \theta) = const.$ in equation \ref{eq:total_potential}). 
However, since the gravitational potential itself depends on the planetary volume, equation \ref{eq:total_potential} has to be solved iteratively for a self-consistent solution. 
This calculation method is implemented in the method "Theory of Figures $4^{\text{th}}$ order" (ToF), that is
derived and explained in \cite[e.g.,][]{Zharkov1970,Zharkov1975,ZharkovVladimirNaumovich1978Popi,Hubbard2014,Nettelmann_2017}. For given planetary radius, mass, rotation period and density profile, ToF $4^{\text{th}}$ order estimates the gravitational coefficients $J_2$, $J_4$, $J_6$, $J_8$, the MoI, as well as the planetary shape (polar radius). Note that only even order $J$ values are considered, as ToF assumes uniform rotation (i.e. the planet to be in HE) and neglects dynamics (e.g. winds) or hemispheric asymmetries that could lead to non-zero odd harmonics. \\

\cite{Helled2011} showed that Uranus' and Neptune's density profile can be represented with a continuous $6^{\text{th}}$-order polynomial. In this work we represent the density profile by (up to) three piece-wise arranged polytropes. This more generous approach additionally allows for up to two density discontinuities ("density jumps"), that can account for potential sharp compositional changes (e.g., rain-out due to immiscible materials). \cite{Neuenschwander_2021} showed for Jupiter, that polynomials and polytropes span a similar solution space and are, in this respect, considered equivalent.
A polytrope connects the pressure $P$ and density in the following way:
\begin{equation}
    P = K\rho^{1+\frac{1}{n}},
    \label{eq:polytrope}
\end{equation}
where $K$ is the polytropic constant and $n$ the polytropic index. 
We use our implementation of ToF to ensure  that the corresponding polytropic relation is met at each point in the fully converged planet.
We call it a "piece-wise" arrangement when different polytropes represent different radial regions of a planet (e.g. core, mantel and envelope region). Three polytropes give a total of 8 free parameters: $n_i$ \& $K_i$ (for $i \in [1,2,3]$), $r_{1,2}$ and $r_{2,3}$. $r_{1,2}$ is the transition radius between the outermost and the middle polytrope and $r_{2,3}$ the transition radius between the middle and the innermost polytrope. 
For simplicity, we refer to the innermost polytrope as the deep interior. \\
More information on the relation between the polytropic indices and the resulting density profile is given in Appendix \ref{subsec:polytropes}. 
It should be noted that the regions represented by the different polytropes do not represent distinct layers of a given (well mixed) composition. Therefore, our parameter space includes more general solutions such as models with composition gradients and could represent interiors with non-adiabatic temperature profiles.  
Since there is no unique solution for the internal structure, we consider a relatively large parameter space. 
We allow the innermost polytrope to extend up to 70\% of the planet's radius and comprise up to 12 M$_\oplus$ of the total planetary mass. 
The maximum central density we consider is $18,000$ kg$\cdot$m$^{-3}$ which is beyond the expected density of rock at the expected core pressure ($\sim 20 \text{ Mbar}$) and core temperature ($\sim 10,000 \text{ K}$) of Uranus and Neptune \citep{sesame7100,Thompson1974,Musella2019}. \\
We use the same method and the same simplex optimization algorithm as described in \cite{Neuenschwander_2021} (also described in \cite{Lagarias1998}, and implemented in MATLAB's 'fminsearch' function). It should be noted that this algorithm does not necessarily cover the entire parameter space of possible solutions but are used to examine a large part of the solution space. 
This allows us to converge to a model that reproduces the measured $J_2$, $J_4$, within their uncertainties (see table \ref{tab:planet_properties}) and $M$, $a$, and $P_{rot}$, exactly. We refer to such a model as a \textit{good solution}. For simplicity, we assume that the uncertainty on the gravitational moment is uniformly distributed, which leads to a relatively large range of possible   solutions.  \\
Although, in principle, the simplex algorithm does not require boundaries in the parameter space, we set the following limits in $K_i$, $n_i$, and $r_{1,2}$: $K_i > 0$ must be strictly positive. This is to prohibit negative pressures and densities. Nevertheless, no upper limit for $K_i$ is introduced. $n_i = (0,2]$ must be within 0 and 2. In the limiting case of $n_i \rightarrow 0$, the polytrope describes an incompressible material (i.e., a material with constant density, independent of pressure). The limit $n = 2$ is set high enough that it is not reached (for Uranus and Neptune $n_i$ values of no larger than 1.2 are obtained, see Figure \ref{fig:polytropic_indices} in section \ref{subsec:polytropes}). Finally and naturally, $r_{1,2}= (r_{2,3},1)$ must be greater than $r_{2,3}$ and smaller than the normalized surface radius of the planet.  \\
In this study, we focus on the relative differences between the inferred solutions. 

Below we use the following parameter space of \textit{planetary models} for Uranus and Neptune. 
Each \textit{planetary model} consists of different planetary characteristics such as the rotation period or the gravity field as described in Table \ref{tab:planet_properties}. The \textit{planetary models} "Uranus--" and "Neptune+" assume $P_{\text{HAS}}$ for the  rotation period, whereas "Uranus" and "Neptune" use the one of $P_{\text{Voy}}$. "Uranus corr dyn" and "Neptune corr dyn" use wind corrected $J_4$ values, as proposed by \cite{Kaspi.2013}. 
The same is true for "Uranus-- corr dyn" and "Neptune+ corr dyn" but using $P_{\text{HAS}}$.
Note that for both Uranus and Neptune we only consider the deepest winds 
with the largest effect on $J_4$, as this acts as an upper boundary. \\
For "Neptune $J_{\text{pre}}$" we use the same uncertainty in $J_2$ and $J_4$ as for Uranus, together with $P_{\text{Voy}}$ 
and investigate how more accurate $J_2$ and $J_4$ values affect the solution space in  $J_6$, $J_8$, and MoI. 

\section{Results}  \label{sec:results}   

\begin{figure*}
    \centering
    \includegraphics[width = 0.5\textwidth]{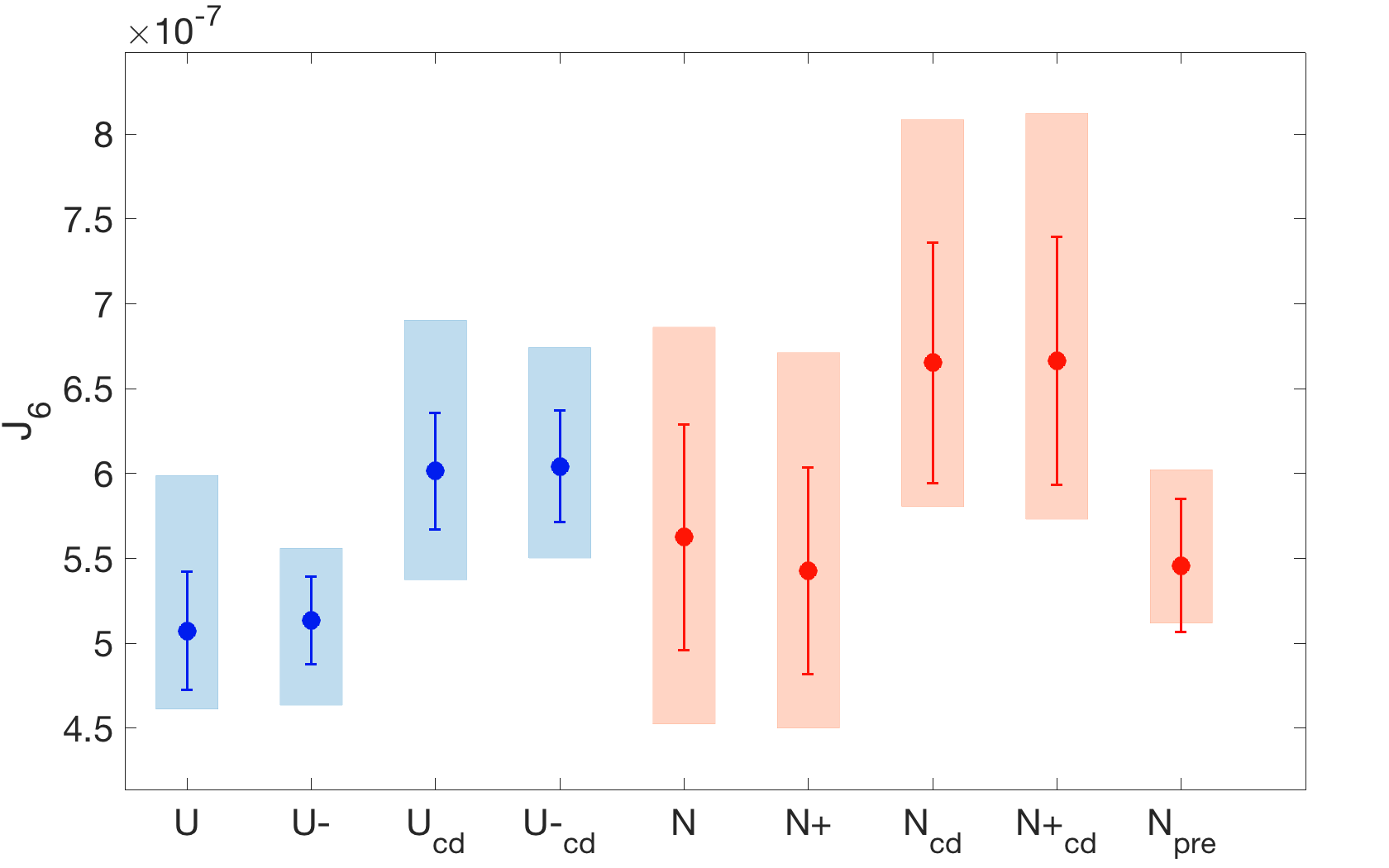}%
    \includegraphics[width = 0.5\textwidth]{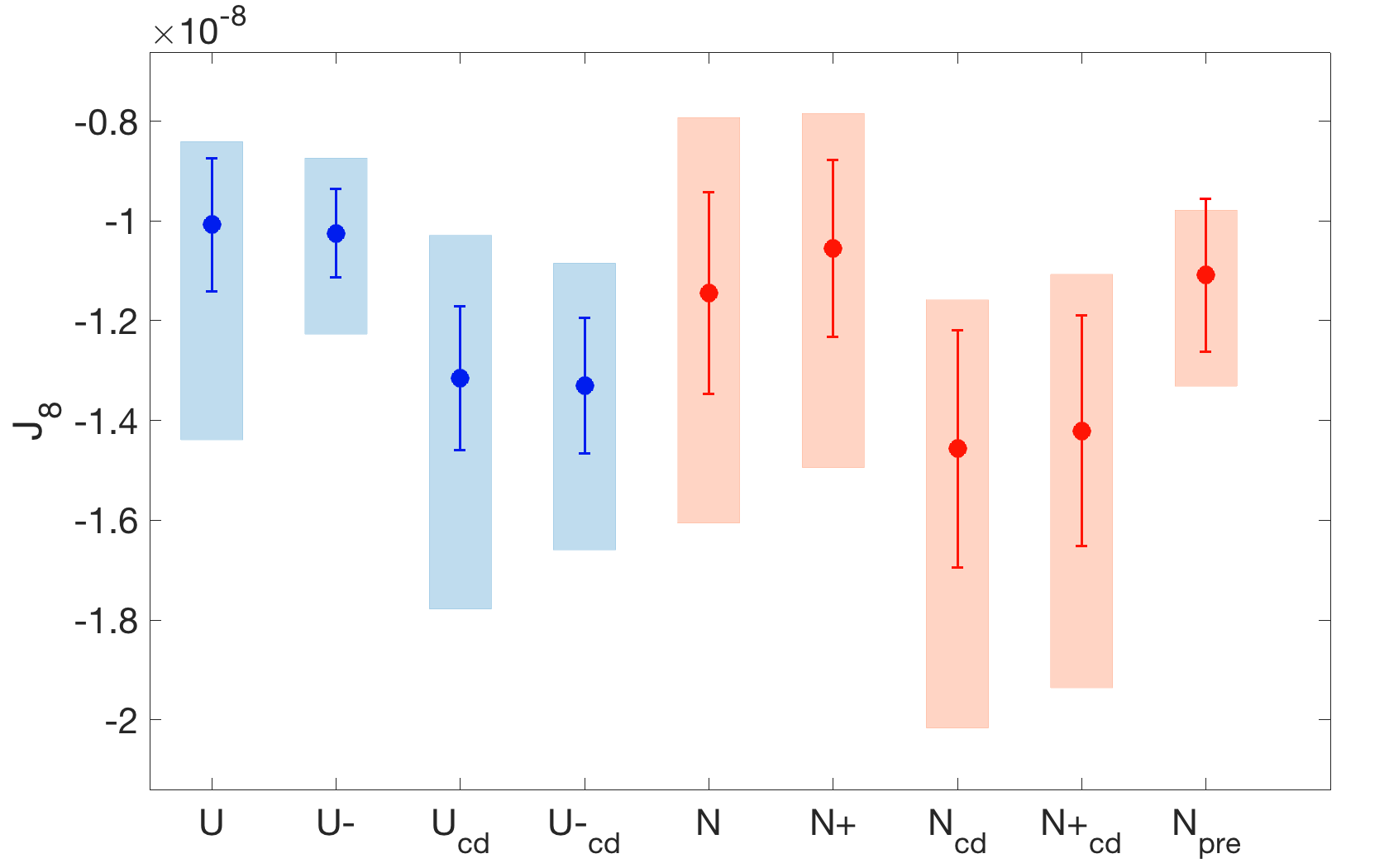}
    \caption{Predictions of $J_6$ (left panel) and $J_8$-values (right panel) for different \textit{planetary models} of Uranus (blue colored) and Neptune (red colored). The \textit{planetary models} are collected along the x-axis with abbreviated names ("U": Uranus, "N": Neptune, "cd": corr dyn).  The dots mark the mean values, the  bars the standard deviation and the boxes show the full solution range.}
    \label{fig:fancy_plots}
\end{figure*}

For all \textit{planetary models} of Uranus and Neptune a broad parameter space has been investigated. Details on the spanned solution space for each \textit{planetary model} can be found in appendix \ref{sec:solution_space}. 
In this Section we present predictions for $J_6$ \& $J_8$ for various \textit{planetary models} in \ref{subsec:J6J8_predictions}. In \ref{subsec: effect of dynamics} we present the effect of different rotation periods and wind characteristics on the solution space. We then demonstrate how improved knowledge of $J_6$ and $J_8$ can  further constrain the depth of the winds in \ref{subsec:constraining_J_values} and how improved knowledge of $J_2$ and $J_4$ further constrains the internal structure in \ref{subsec:better_known_J2_J4}. In \ref{subsec:constraining_power_of_shape} we investigate the relation between the planetary shape and its rotation period. Finally, we demonstrate that an accurate determination of the MoI can constrain the dynamics of Uranus and Neptune in \ref{sec:The MoI}.

\subsection{$J_6$ \& $J_8$ predictions} \label{subsec:J6J8_predictions}
For both Uranus and Neptune, only the low order gravitational coefficients $J_2$ and $J_4$ are known. Here we use the \textit{planetary models} to predict $J_6$ and $J_8$ of Uranus and Neptune. 
The results are shown in Figure \ref{fig:fancy_plots}. The left (right) panel shows the predicted range of $J_6$ ($J_8$) values for Uranus (blue colored) and Neptune (red colored). The dots mark the corresponding mean, the error bars the standard deviation, and the colored boxes the full value range of each \textit{planetary model} for Uranus and Neptune. The predicted ranges of $J_6$ and $J_8$ are listed in Table \ref{tab:results}. \\
We find that the predicted $J_6$ and $J_8$ value ranges of Uranus and Neptune are very similar, which is expected given that they both have very similar values of $J_2$ and $J_4$. In addition, the uncertainties are smaller for Uranus in comparison to Neptune due to the more accurate determinations of $J_2$ and $J_4$. 
For both planets, the predicted values of $J_6$ and $J_8$ strongly depend on the assumed planetary dynamics.
We note that different assumed uniform rotation periods do not significantly  affect the $J_6$ and $J_8$  predictions, except for Uranus, where a faster rotation period slightly constrains the $J_6$ and $J_8$ predictions.
However, dynamically corrected $J_4$ values change significantly the predicted values in $J_6$ and $J_8$. For both planets, deep winds decrease (increase) the mean value of $J_6$ ($J_8$) by roughly 20\% (30\%). \\
We conclude that it is crucial to constrain the depth of the winds in Uranus and Neptune to further constrain the predictions of $J_6$ and $J_8$. At the same time, accurate measurements of the higher order harmonics can be used to constrain the depth of the winds (see \ref{subsec: effect of dynamics}). To confirm wind depths of the order of $\sim$1100 km  for Uranus and Neptune, relative uncertainties in $J_6$ and $J_8$ better than 10\% (preferably a few percent) are required. For shallower winds an increased precision in $J_6$ and $J_8$ is necessary. This is because the solution spaces in $J_6$ and $J_8$ of wind-corrected and non-wind-corrected \textit{planetary models} converge. Also the required accuracy in $J_6$ and $J_8$ are  different for Uranus and Neptune since currently Uranus has more accurate $J_2$ and $J_4$ determination. As a result, to have the same constraining power, measurements of Uranus' $J_6$ and $J_8$ values have to be $\sim$40\% more accurate than for Neptune. \\
Finally, we find that a better measurement of $J_2$ and $J_4$, as implemented in the \textit{planetary model} "Neptune $J_{\text{pre}}$" would significantly constrain the predicted $J_6$ and $J_8$ values. The parameter space of solutions for "Neptune $J_{\text{pre}}$" then becomes comparable to that of "Uranus".

\subsection{Effect of atmosphere dynamics} \label{subsec: effect of dynamics}

\begin{figure*}
    \centering
    \includegraphics[width = 0.5\textwidth]{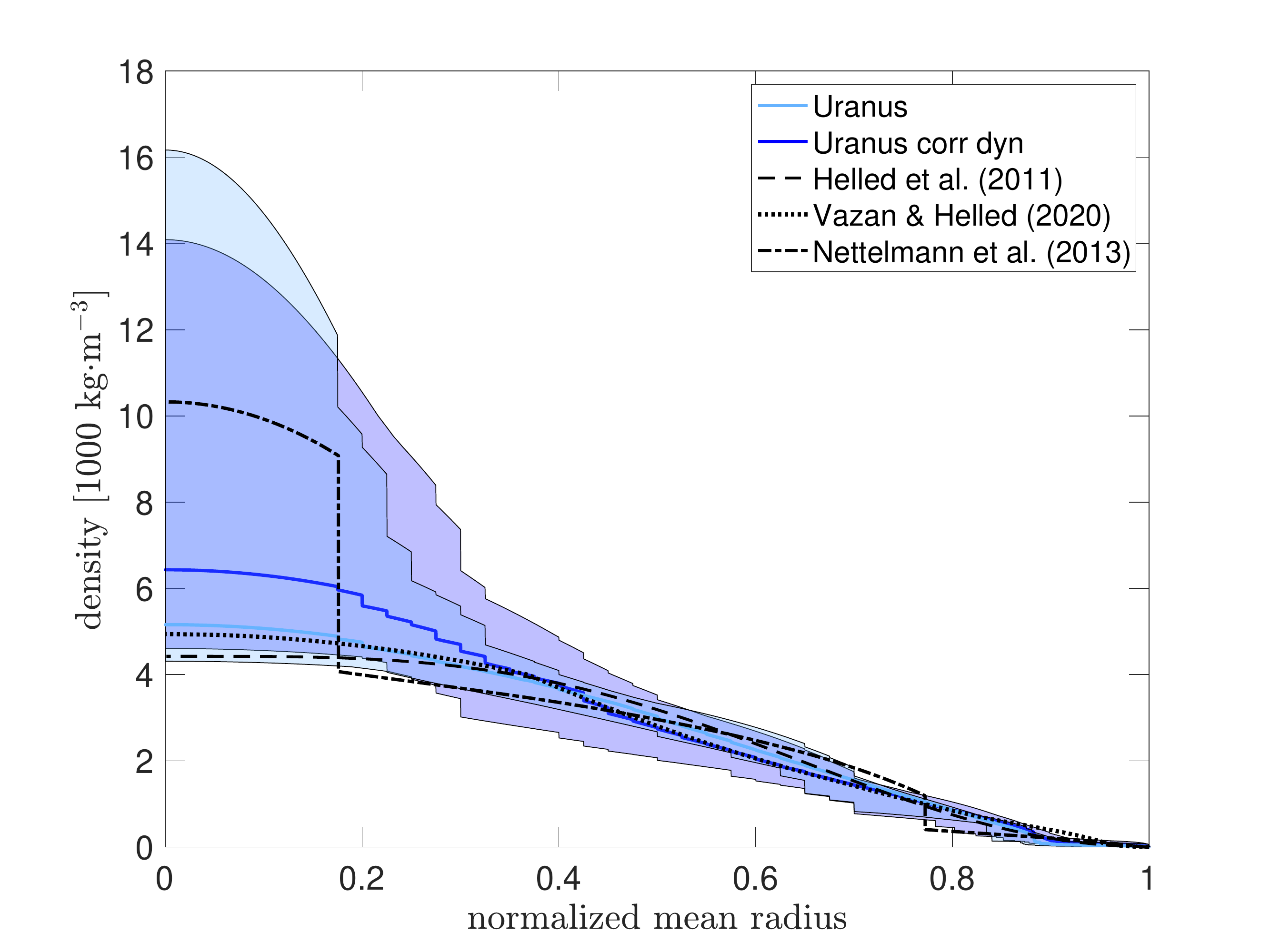}%
    \includegraphics[width = 0.5\textwidth]{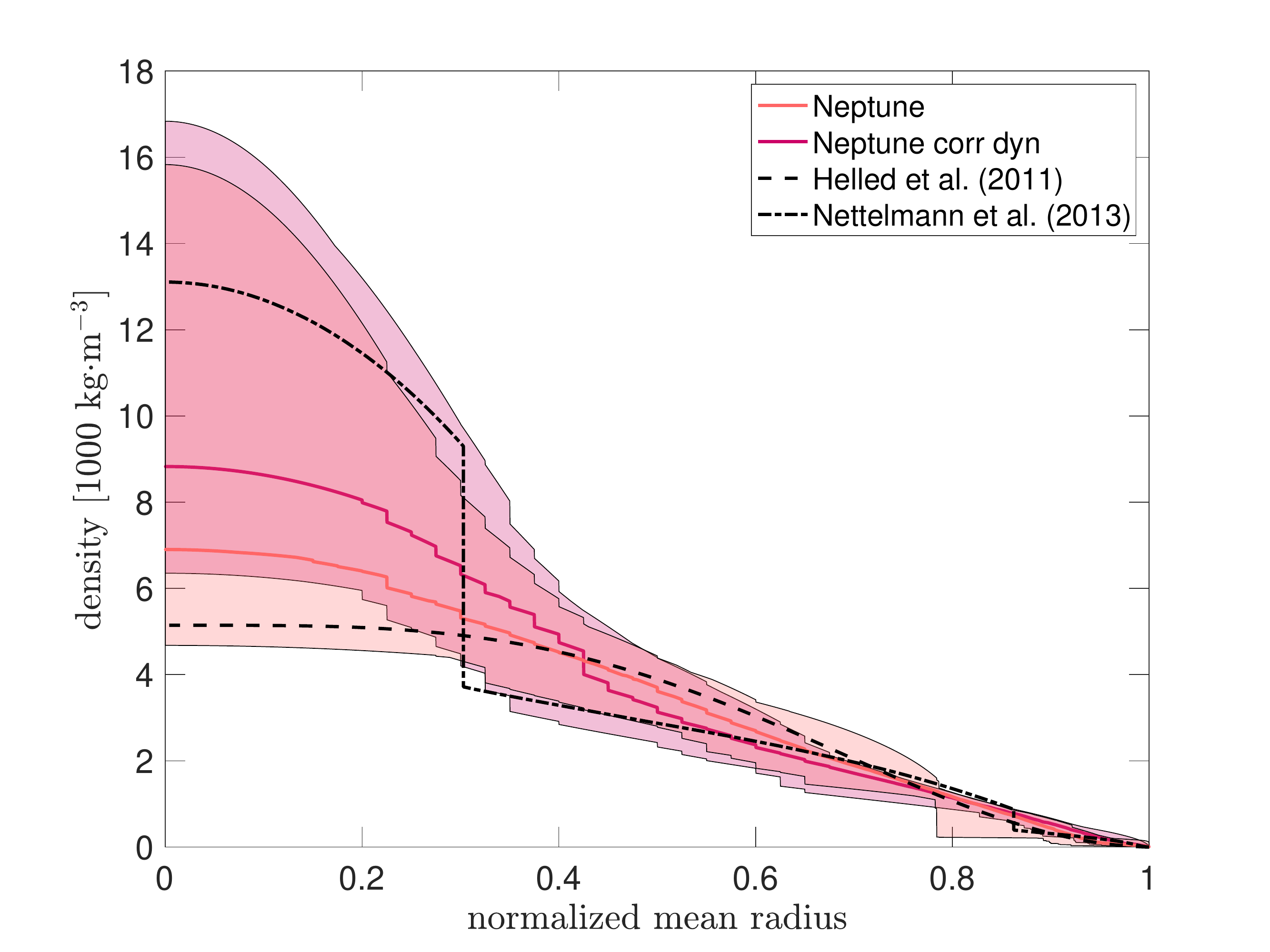}
    \includegraphics[width = 0.5\textwidth]{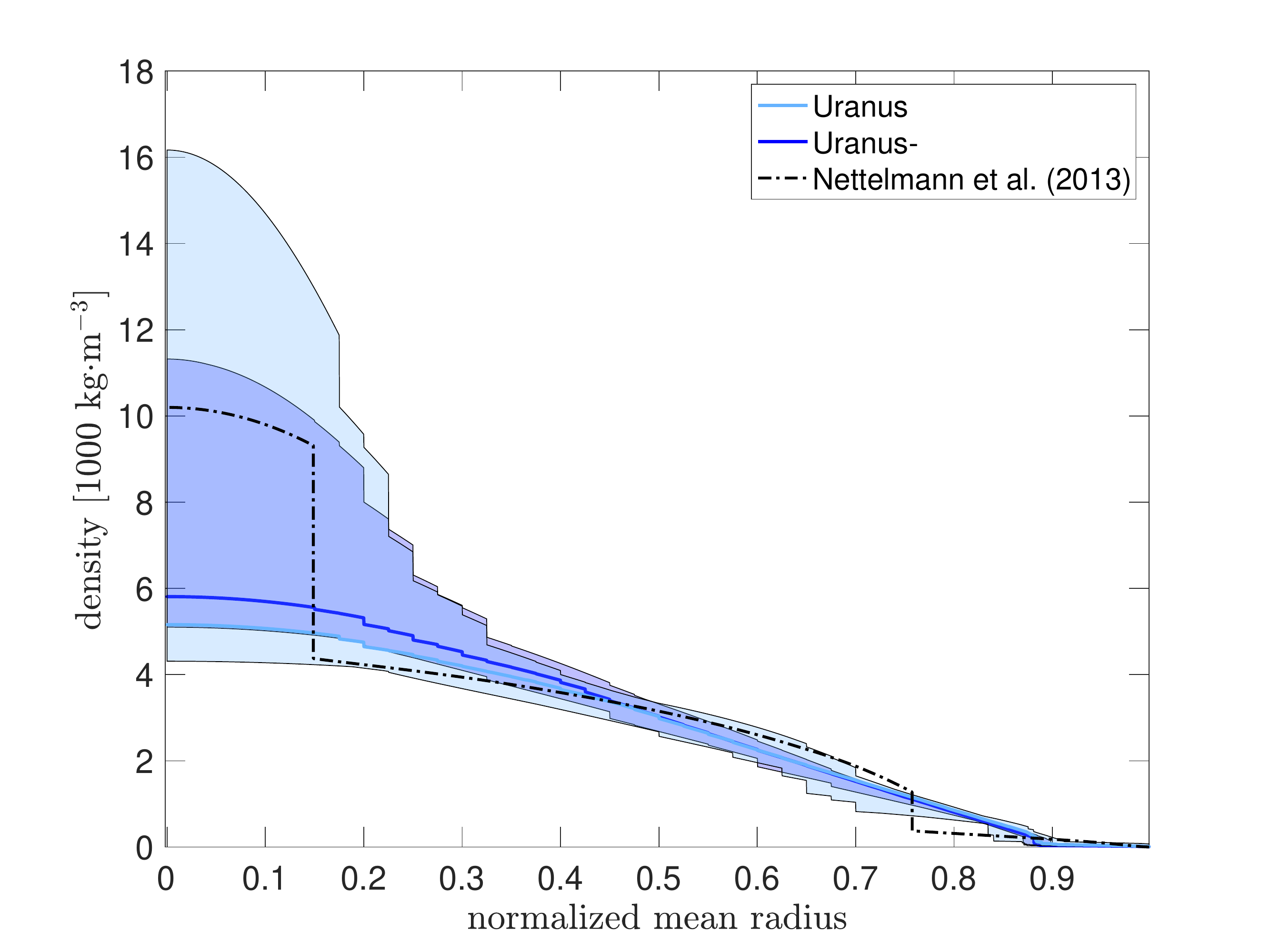}%
    \includegraphics[width = 0.5\textwidth]{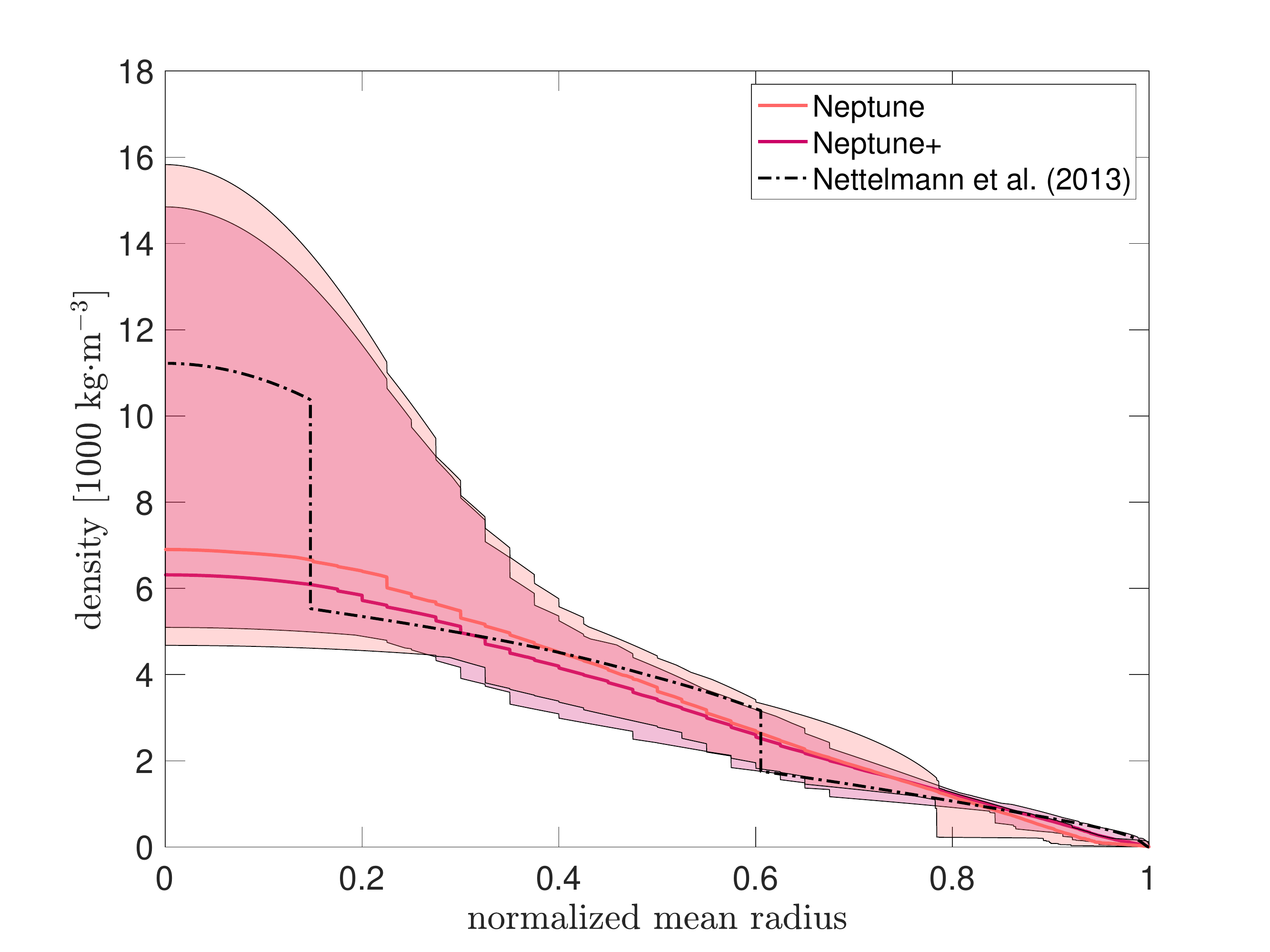}
    \caption{Inferred density profiles for Uranus (left column) and Neptune (right column) with different assumed rotation periods (lower panels) and depths of the winds (upper panels). The solid curves represent the sample median, whereas the shaded area incorporate 96\% of all solutions. In the upper panels, we include solutions for Uranus and Neptune of \protect\cite{Helled2011}, \protect\cite{NETTELMANN2013} and \protect\cite{Vazan2020} (black dashed, dash-dotted and dotted lines, respectively). In the lower panels we include for "Uranus--" and "Neptune+" density profiles of \protect\cite{NETTELMANN2013} (black dash-dotted curves).   }
    \label{fig:density_plots_dynamics}
\end{figure*}
Below we investigate the effect of the depth of the winds on the inferred $J_6$ and $J_8$ values, MoI, polar radius, and the density profiles. We compare the results from "Uranus" and "Neptune" (based on $J_{4}^{meas}$) with "Uranus corr dyn" and "Neptune corr dyn", respectively. \\
The upper left panel of Figure \ref{fig:density_plots_dynamics} shows the solution space of internal structure models of "Uranus" (light-blue themed) and "Uranus corr dyn" (dark-blue themed). The colored curves mark the corresponding sample median and the colored shaded areas comprises 96\% of all solutions.  
The upper right panel is similar but presents the results for "Neptune" (light-red colored) and "Neptune corr dyn" (dark-red colored). 
For comparison, published density profiles of Uranus and Neptune from \cite{Helled2011}, \cite{Vazan2020} and \cite{NETTELMANN2013} are included. These published density profiles correspond to a subset of our solutions. 

We find that atmosphere dynamics can have a significant impact on the density profiles: for both Uranus and Neptune solutions based on wind-corrected $J_4$ tend to have a higher central density in comparison to the solutions without wind corrections. 
Their mean central density is $\sim$25\% higher for both, "Uranus corr dyn" and "Neptune corr dyn". This also results in higher central pressures. 
As a result, solutions that include the wind correction on $J_4$ tend to have more massive innermost polytropes for a given $r_{2,3}$ or a smaller $r_{2,3}$ for a given mass of the innermost polytrope, respectively (see Figure \ref{fig:solution_space} in appendix \ref{sec:solution_space} for more details). Since the total planetary mass is fixed, more centrally condensed material leads to a less massive intermediate region between a normalized radius of 0.4 and 0.8 (for both Uranus and Neptune), where-after the density has to increase again compared to the models without dynamical corrections in $J_4$. This is necessary to fit the higher $|J_4|$ values (see table \ref{tab:planet_properties}). 
As a result, the MoI values are smaller for solutions based on the wind-corrected gravity field (see Table \ref{tab:results}). Further, the predicted $J_6$ and $J_8$ values get larger and smaller, respectively. This is also  expected since the winds of Uranus and Neptune are concentrated to the outermost region of the planets, particularly affecting the higher-order $J$ values \citep[e.g.,][]{Hubbard1999,Kaspi_2010}.
Interestingly, the density profiles with the wind correction of $J_4$ allow for a broader solution space at $r \sim 0.2-0.65$ for Uranus and $r \sim 0.3-0.5$ for Neptune but are more constrained in Neptune's outer region ($r> 0.65$). \\
Although the shape of a planet is strongly correlated with $J_2$ (see Figure \ref{fig:J2_polar_radius} in appendix \ref{section:polar_radius}) and not necessarily with $J_4$, for Uranus with wind corrected $J_4$ values, polar radii are found to be slightly higher (Table \ref{tab:results}).

\begin{table*}
\centering
 \begin{tabular}{l| l c l c}
 \hline 
\multicolumn{1}{c}{\textit{planetary model}} & \multicolumn{1}{c}{$J_6$ [$\times10^{-8}$]} & \multicolumn{1}{c}{$-J_8$ [$\times10^{-9}$]} & \multicolumn{1}{c}{MoI}  & \multicolumn{1}{c}{polar radius [km]} \\
 \hline 
Uranus                      & $46.12-59.90$   & $8.42-14.39$    & $0.22594-0.22670$  & 25,052.7 \\
Uranus--                    & $46.36-55.60$   & $8.74-12.27$    & $0.21919-0.21964$  & 25,022.5 \\
Uranus corr dyn             & $53.76-69.04$   & $10.29-17.78$   & $0.22529-0.22625$  & 25,052.8 \\
Uranus-- corr dyn           & $55.03-67.44$   & $10.85-16.59$   & $0.21840-0.21920$  & 25,022.6 \\ 
Neptune                     & $45.26-68.63$   & $7.93-16.05$    & $0.23727-0.23900$  & 24,315.9 \\
Neptune+                    & $45.02-67.12$   & $7.85-14.94$    & $0.25248-0.25431$  & 24,382.9 \\
Neptune corr dyn            & $58.05-80.85$   & $11.58-20.16$   & $0.23547-0.23826$  & 24,316.0 \\
Neptune+ corr dyn           & $57.32-81.21$   & $11.07-19.36$   & $0.25007-0.25355$  & 24,382.9 \\
Neptune $J_{\text{pre}}$    & $51.20-60.24$   & $9.78-13.31$    & $0.23808-0.23860$  & 24,315.9 \\
\hline 
\end{tabular}%

\caption{Predicted values of $J_6$, $J_8$, the MoI, and the polar radius of all investigated \textit{planetary models} of Uranus and Neptune. The standard deviation in polar radii is is smaller than a few 100 meters.}%
\label{tab:results}%
\end{table*}

\subsubsection{Assumed uniform rotation period} \label{sec:diff_rot_rate}
The rotation periods of Uranus and Neptune are not well determined. Theoretical estimates suggest modified rotation periods that differ from the Voyager periods by 40 min for Uranus and 1 h 20 min for Neptune \citep[e.g.,][]{Helled2010_prot}. Here, we investigate the effect of different assumed uniform  rotation periods on the inferred density profile solution space, the MoI, the gravity field, and the planet's shape. We assume two rotation periods for  Uranus and Neptune: $P_{\text{Voy}}$ and $P_{\text{HAS}}$. \\

The assumed uniform rotation period affects the planetary  centrifugal potential. Therefore, a faster rotation period leads to a more oblate planet. For Uranus with  $P_{\text{HAS}}$ we notice a decrease in the polar radius $b_0$ of 0.12\% (30.2 km), compared to the Voyager rotation period. For Neptune with $P_{\text{HAS}}$ we observe an increase of 0.27\% (67 km) in $b_0$ in comparison to $P_{\text{Voy}}$ (see Table \ref{tab:results}). These results are consistent with the results of  \cite{Helled2010_prot}. 
Figure \ref{fig:density_plots_dynamics} shows the density profile solution space of Uranus (lower left panel) and Neptune (lower right panel) assuming  $P_{\text{Voy}}$ (light-blue and light-red themed, respectively) and $P_{\text{HAS}}$ (dark-blue and dark-red themed, respectively). The solid curve marks the corresponding sample median, whereas the shaded areas include 96\% of all solutions. The black dash-dotted curves mark solutions of \cite{NETTELMANN2013} for "Uranus--" and "Neptune+". We find that while for "Neptune+" the solution of \cite{NETTELMANN2013} is embedded in our density solution space, this is not entirely true for "Uranus--". For "Uranus--" our local optimization algorithm could not generate a large enough density discontinuity at
the transition radius $r_{1,2}$. 
Nevertheless, we still observe a large variety of density profiles in the deep interior. Additionally, we are mainly interested in relative changes 
between different \textit{planetary models} and do not need a complete solution space. 
Our findings, therefore, are expected to persist even when considering a larger parameter space. \\
We find that faster (slower) rotation period leads to 13\% higher (9\% lower) mean central densities and lower (higher) densities in the outer envelope region ($r>0.6$, $r>0.75$) for Uranus and Neptune, respectively. This is expected because a faster rotation period increases the centrifugal force, which in turn pushes mass into the outer region. Since, $J_2$ and $J_4$ are unchanged, the pushed-out mass has to be light (low density), which decreases the density in the outer envelope. This inevitably leads to a higher density in the deep interior, as the total mass has to be conserved.  \\
A more centrally condensed interior also leads to a slightly smaller mean $J_4$-value (although still within its measurement uncertainty). This in turn also affects the $J_6$ and $J_8$ solution range, as $J_4$, $J_6$ and $J_8$ are correlated (see Figure \ref{fig:fancy_plots} \& Figure \ref{fig:J4_vs_J6_vs_J8}). The opposite behavior is true for a slower rotation period. \\
Finally, a more centrally condensed interior also leads to smaller MoI values. Therefore, a faster rotating planet tends to have a lower MoI. 
A detailed analysis of the dependency of the MoI  on the rotation period is presented in section \ref{sec:The MoI}. 
In summary, we conclude that the planetary rotation period has a major impact on the shape (polar radius $b_0$), the density distribution, and the MoI value. \\
Accurate measurements of the planetary shape, in particular the polar radius as well as a determination of the MoI could further constrain the planetary rotation period of Uranus and Neptune.

\subsection{Relation between the $J$ values, the planetary shape, and the depth of the winds} \label{subsec:constraining_J_values}
In this section we investigate whether more accurate determinations  of the gravity fields of Uranus and Neptune could be used to  constrain the depth of the winds. We also investigate the relation between the planetary shape, the MoI and $J_2$. \\
Figure \ref{fig:fancy_plots} and Table \ref{tab:results} show that for Uranus $J_6=(59.9-69.04)\cdot10^{-8}$ and $J_8=(14.39-17.78)\cdot10^{-8}$ values can only be explained by the existence of deep winds with a penetration depth of more than 250 km. For Neptune the same is true for $J_6=(68.63-80.85)\cdot10^{-8}$ and $J_8=(16.05-20.16)\cdot10^{-8}$. On the other hand, values of $J_6 =(46.12-53.76)\cdot10^{-8}$ and $J_8=(8.42-10.29)\cdot10^{-8}$ do not allow for deep winds (penetration depth $\sim1,100$ km) in Uranus. In Neptune deep winds are forbidden for $J_6=(45.26-58.05)\cdot10^{-8}$ and $J_8= (7.93-11.58)\cdot10^{-8}$.

\begin{figure}
    \centering
    \includegraphics[width = 0.5\textwidth]{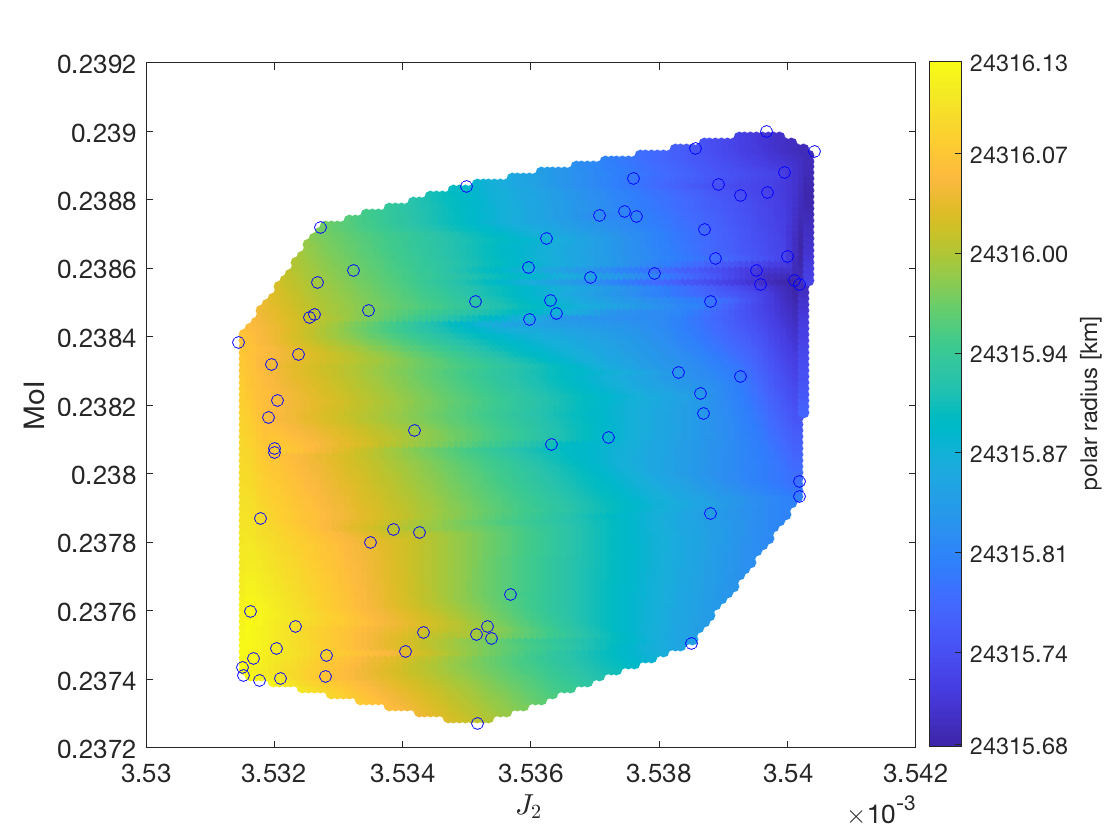}
    \caption{The relation between Neptune's polar radius, MoI and $J_2$. Blue circles mark \textit{good solutions}. The colors correspond to the inferred polar radius that has been interpolated between the solutions. 
    }
    \label{fig:Neptune_polar_radius}
\end{figure}

Therefore accurate measurements of either $J_6$ (with a relative uncertainty of 0.1) or $J_8$ (with a relative uncertainty of 0.1), for both planets, could constrain the depth of the winds \citep[e.g.,][]{Hubbard1999,Kaspi_2010}. \\
It is known that $J_2$ strongly correlates with the planetary shape \citep[e.g.,][]{Mecheri2004, Helled2011} and our results confirm this correlation (see Figure \ref{fig:J2_polar_radius} in appendix \ref{section:polar_radius}). This implies that an accurate measurement of $J_2$ can further constrain Neptune's polar radius and vice-versa. 
Figure \ref{fig:Neptune_polar_radius} shows the relation between $J_2$ (x-axis), MoI (y-axis) and polar radius (color coded) for Neptune. We observe a clear trend from low $J_2$ ($< 3.5325\cdot10^{-3}$) and MoI ($< 0.2378$) to high MoI ($> 0.2384$) and $J_2$ values ($> 3.5385\cdot10^{-3}$). 
We conclude that a measurement of Neptune's polar radius cannot only be used to constrain $J_2$  but also the MoI value: a MoI value smaller than MoI $< 0.23760$ is related to a polar radius larger than $b_0$ $> 24,316.11$ km and a MoI larger than $> 0.23855$ is related to a polar radius smaller than $b_0$ $< 24,315.71$ km. We clearly show that there is no one-to-one correspondence between the MoI and $J_2$, confirming that the Radau-Darwin relation is only a rough approximation \citep[e.g.,][]{Podolak2012,GAO2013,Neuenschwander_2021}.  

\subsection{The importance of accurate measurements of $J_2$ and $J_4$} \label{subsec:better_known_J2_J4}
Roughly speaking, better knowledge of the $J$ values (e.g. smaller uncertainty and/or measurement of higher order harmonics) further constrains the  solution space.  However, this might not be entirely true as the $J$ values are "blind" (not sensitive) to masses close to the planet's centre.
Therefore, an ambiguity in the deep interior region is always expected to persist for models based solely on gravity data. 

For Uranus, both, $J_2$ and $J_4$ are estimated more accurately than for Neptune (see Table \ref{tab:planet_properties}). As a result, the solution range in $J_6$, $J_8$, polar radii, MoI are more constrained for Uranus. This is shown in Table \ref{tab:results} and Figure \ref{fig:fancy_plots}. \\
Here, we investigate how an improved measurement of Neptune's gravity field, comparable to the uncertainty in Uranus' measurement, can constrain its predicted $J_6$ and $J_8$ as well as its MoI value.  

We therefore consider a \textit{planetary model} for Neptune with artificially improved precision in $J_2$ and $J_4$ ("Neptune $J_{\text{pre}}$"). 
We then compare this \textit{planetary model} with "Neptune", focusing on the predicted  $J_6$ and $J_8$ values, the polar radius, and the MoI. 
For "Neptune $J_{\text{pre}}$" we use the same uncertainty in $J_2$ and $J_4$ as estimated in "Uranus" and apply it around the estimated mean values of \cite{Jacobson.2009}. Therefore, in this setup, the underlying assumption is that the true values of Neptune's $J_2$ and $J_4$  are near the median of the measurement. This is of course not necessarily the case, and the real $J_2$ and $J_4$ values could be anywhere within the measurement uncertainty (see appendix \ref{sec:improvedJs} for discussion).  
Compared to the previous estimates of \cite{Jacobson.2009}, "Neptune $J_{\text{pre}}$" has decreased uncertainties in $J_2$ by $\approx 85\%$ and in $J_4$ by $\approx 75\%$. \\
We find that the artificially improved uncertainty in $J_2$ and $J_4$ (in "Neptune $J_{\text{pre}}$") is significant and further constrains the solution range in $J_6$, $J_8$, and the MoI (Figure \ref{fig:fancy_plots} and Table \ref{tab:results}). 
We find that the standard deviation in $J_6$ decreases by 41\%, in $J_8$ by 24\% and in the MoI by 55\%. 
Similarly, a more accurate measurement of Uranus' gravity field will further constrain its internal structure. 
We therefore conclude that constraining the $J_2$ and $J_4$ values for Uranus and Neptune is highly desirable. 
We also suggest that measurements of the gravitational fields of Uranus and Neptune by future missions are desirable.   

\subsection{Constraining power of the planetary shape} \label{subsec:constraining_power_of_shape}
\begin{figure}
    \centering
    \includegraphics[width = 0.5\textwidth]{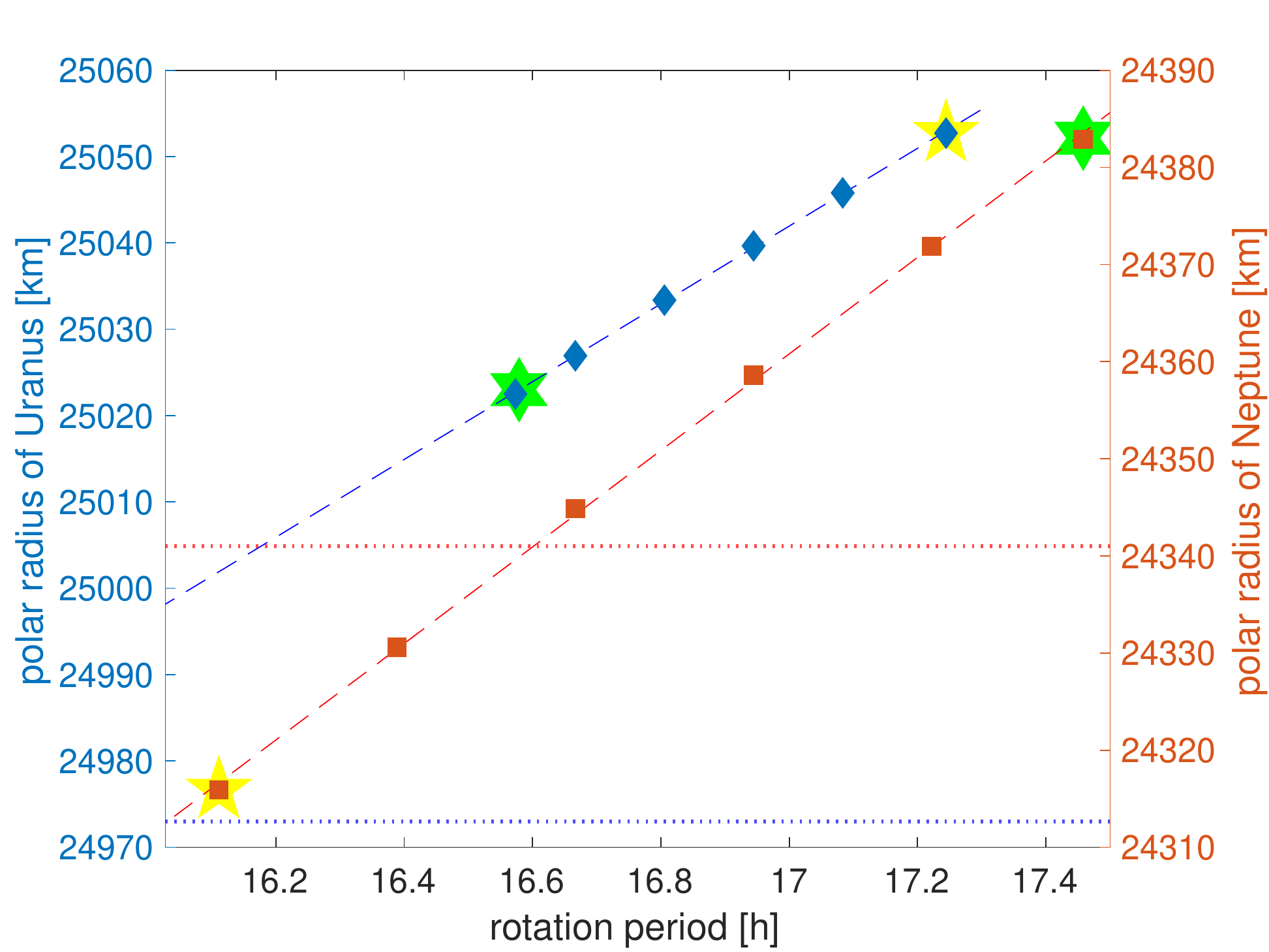}
    \caption{The relation between the polar radius and the assumed (uniform) rotation period for Uranus (blue diamonds) and Neptune (red squares). The dashed lines mark the best-fitting curve.  
    The dotted lines mark the estimated polar radius of 24,973 km (Uranus) and 24,341 km (Neptune). Rotation periods belonging to $P_{\text{Voy}}$ are highlighted with a yellow star. 
    Results from \protect\cite{Helled2010_prot} (for rigid body rotations) based on $P_{\text{HAS}}$ are represented by green pentagrams.
    } 
    \label{fig:different_rot_uran_nept}
\end{figure}

The planetary shape strongly depends on the rotation period. In a faster rotating planet, material gets pushed outside perpendicular to the planet's rotation axis. This results in a more oblate shape. To first order, the planetary flattening can be used to infer the rotation period.
Figure \ref{fig:different_rot_uran_nept} shows the relation between the polar radius and different uniform rotation periods for Uranus (blue themed) and Neptune (red themed) between $P_{\text{Voy}}$ and $P_{\text{HAS}}$. The blue and red dashed lines mark the best fitting curves for Uranus and Neptune, respectively. Yellow stars highlight the solutions based on $P_{\text{Voy}}$, whereas green pentagrams show solutions of \cite{Helled2010_prot} based on $P_{\text{HAS}}$. For $P_{\text{HAS}}$ our results agree very well with the results of \cite{Helled2010_prot}. The dotted horizontal lines correspond to the estimated polar radius of Uranus (24,973 km), and Neptune (24,341 km), taken from \cite{Archinal2018}.These radii are not measured but obtained by extrapolating the measured radio occultation radius towards the pole. The extrapolation was obtained with the integration of the observed gravity field and the zonal wind velocities \citep[see][for details]{Lindal1985}.
The predicted value of Neptune's equatorial radius depends on the assumed uniform rotation period (see Table \ref{tab:planet_properties} and \cite{Helled2010_prot}). We therefore adapted the equatorial radius for each assumed rotation period.  \\
We notice that Neptune's estimated polar radius corresponds to a rotation period of 16.6015 hours. This is at odds with the Voyager II measurement by nearly 30 min and with $P_{\text{HAS}}$ by $\sim$ 50 min. 
To infer Uranus' estimated polar radius of 24,973~km, a rotation period of 15.625 h is required. This rotation period is shorter than $P_{\text{Voy}}$ and $P_{\text{HAS}}$ by $\sim$ 1 h and 1.5 h, respectively. \\ 
We conclude that there is a clear mismatch between the measured shapes of Uranus and Neptune and the Voyager rotation periods. This may imply that uniform rotation is not applicable for these planets, or that the shapes are significantly modified by the winds \citep[][]{Helled2010_prot}. 
\\
It is clear that robust estimates of the rotation periods (and profiles) as well as the shapes of Uranus and Neptune are desirable.  If one is given, the other can be better estimated. 
Our findings are in agreement with the work of \cite{Helled2010_prot}, where it was shown that for both Uranus and Neptune $P_{\text{Voy}}$ is inconsistent with the inferred shape information from \cite{Lindal1987} and \cite{Lindal1992}. 

\subsection{The importance of MoI} \label{sec:The MoI}

\begin{figure}
    \centering
    \includegraphics[width = 0.5\textwidth]{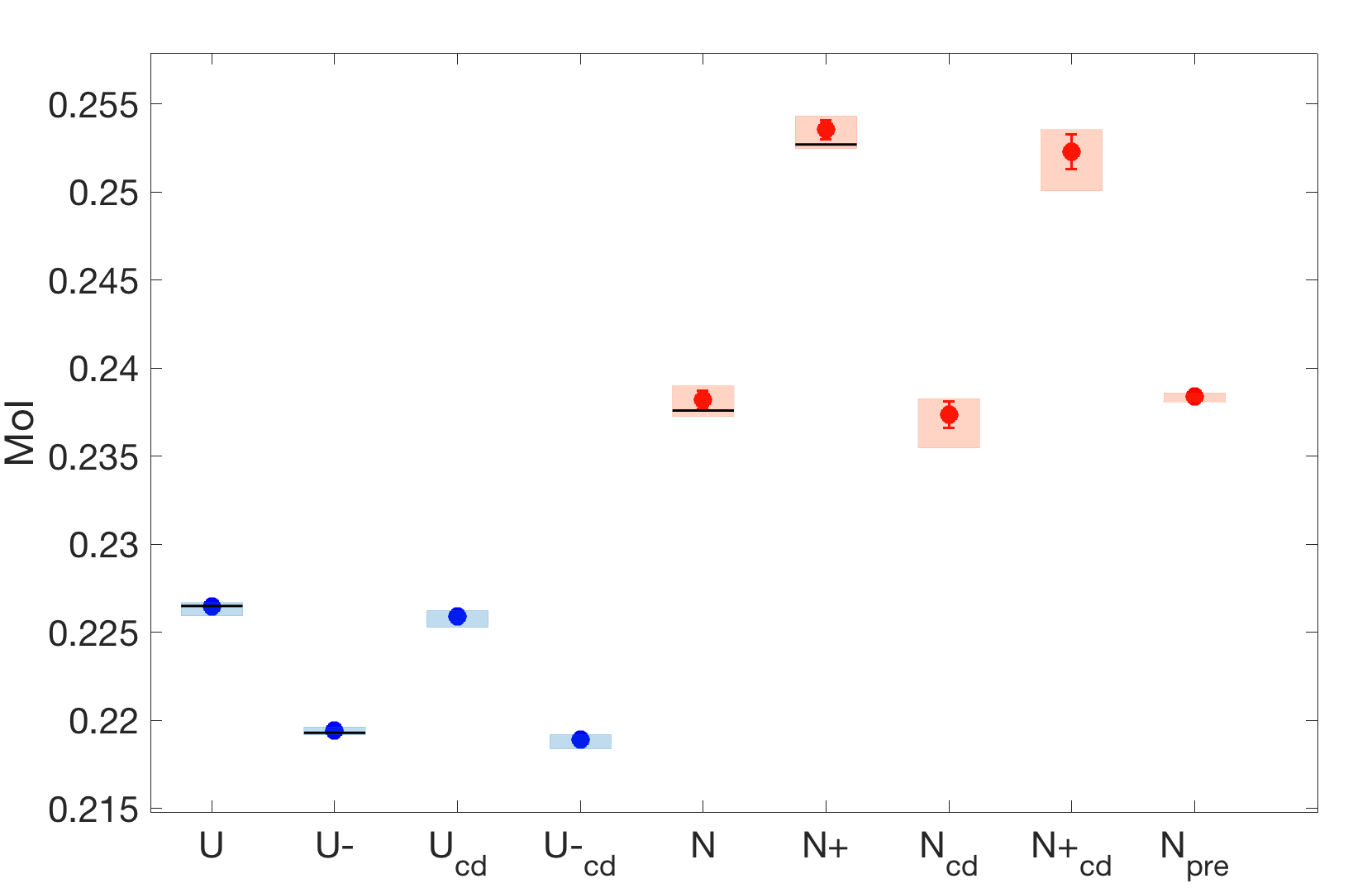}%
    \caption{Inferred MoI values for Uranus (blue colored) and Neptune (red colored). The \textit{planetary models} are collected along the x-axis with abbreviated names ("U": Uranus, "N": Neptune, "cd": corr dyn). Dots mark the mean values, bars the standard deviation and boxes show the full solution range. 
    For selected \textit{planetary models} the black horizontal lines represent solutions from \protect\cite{NETTELMANN2013}.
    }
    \label{fig:fancy_plots_MoI}
\end{figure}
\begin{figure}
    \centering
    \includegraphics[width = 0.5\textwidth]{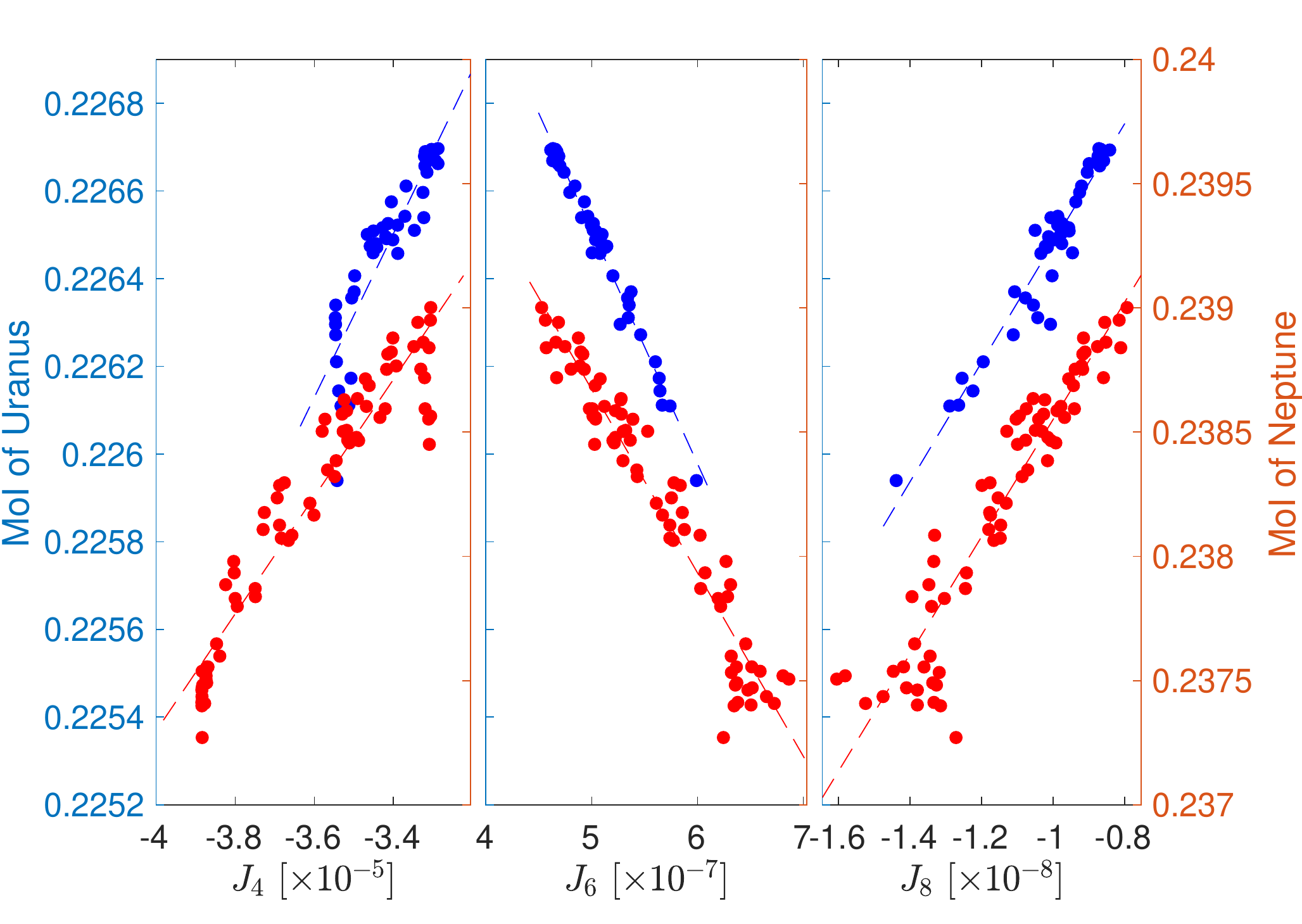}
    \caption{The relation between the MoI and $J_4$ (left panel), $J_6$ (middle panel) and $J_8$ (right panel) for Uranus (blue themed) and Neptune (red themed). The dashed lines mark the best fitting curves.}
    \label{fig:MoI_vs_J4}
\end{figure}

Figure \ref{fig:fancy_plots_MoI} shows the MoI ranges for Uranus (blue) and Neptune (red) for various \textit{planetary models}. The circle marks the mean value, the error bar the standard deviation and the box the whole solution range.
Sometimes the uncertainty is covered by the circle marking the mean value. \\
All the presented MoI values are normalized to the equatorial radius, i.e., MoI $=I/(M\cdot a)$, where $I$ is the axial moment of inertia. When normalized to the planetary equatorial radius, the results of \cite{NETTELMANN2013} are in excellent agreement with our MoI predictions. For illustration, the MoI values of published models of \cite{NETTELMANN2013} are indicated as black horizontal lines in Figure \ref{fig:fancy_plots_MoI}. \\
We find that the MoI value range strongly depends on the planetary rotation period. For a faster rotating Uranus ("U--" and "U--$_{\text{cd}}$") the mean MoI decreases from $0.22647 \xrightarrow{}0.21943$ (a decrease of 3.2\%) and for a slower rotating Neptune ("N+" and "N+$_{\text{cd}}$") the MoI increases from $0.23821 \xrightarrow{}0.25354$ (an increase of 6\%). \\
For rotation periods between $P_{\text{Voy}}$ and $P_{\text{HAS}}$, we expect the inferred MoI values to be within our reported ranges. \\
Additionally, \textit{planetary models} with a wind corrected $J_4$ value ("U$_{\text{cd}}$", "U--$_{\text{cd}}$", "N$_{\text{cd}}$" and "N+$_{\text{cd}}$") experience a minor shift in the MoI range (see section \ref{subsec: effect of dynamics} and Table \ref{tab:results}).
We conclude that the rotation periods of Uranus and Neptune could be further constrained by a measurement of the MoI with a relative uncertainty of $\sim1\%$. 
Additionally, with an independently measured MoI with a relative uncertainty of $\sim0.1\%$, the depths of the winds of Uranus and Neptune can be further constrained. We find that for Uranus (Neptune) MoI values of $0.22529-0.22594$ ($0.23547-0.23727$) can only be explained by winds that  penetrate to depths deeper than $\sim 250$ km. On the other hand, MoI values of $0.22625-0.2267$ (Uranus) and $0.23826-0.239$ (Neptune) would exclude winds with penetration depths of $\sim$ 1,100 km. \\
It should be noted, however, that an accurate  measurement of the MoI is not an easy task and it can  only be obtained with a future space mission that is  designed for this measurement. \\
Naturally, the MoI value can be constrained by flipping the dependencies: an accurate determination of the depth of the winds, or a robust determination of the rotation periods can also constrain the MoI value.
Although there is no one-to-one correspondence between $J_2$ and the MoI (see section \ref{subsec:constraining_J_values}), there is a rather strong correlation between the higher order gravitational coefficients $J_4$, $J_6$, $J_8$, and the MoI. \\
Figure \ref{fig:MoI_vs_J4} shows the relation between the MoI and $J_4$ (left panel), $J_6$ (middle panel) and $J_8$ (right panel) for Uranus (blue dots, corresponding to the y-axis on the left) and Neptune (red dots, corresponding to the y-axis on the right). The colored dashed lines mark the best-fitting curves.  

This behavior is expected as both the higher-order gravitational coefficients $J_4$, $J_6$, $J_8$ and the MoI can be used to constrain the depth of the winds. 
Nevertheless, these inferred relations allow us to further constrain $J_4$, $J_6$ and $J_8$, given a very accurate measurement of the MoI, or to further constrain the MoI with an accurate measurement of the higher order harmonics. 

\section{Summary and Conclusions} \label{sec:summary_and_conclusion}   

In this work we present empirical structure models of Uranus and Neptune where the density profile is represented by (up to three) polytropes. We use these models to predict their $J_6$ and $J_8$ values and to investigate the effect of deep winds and assumed rotations period on the inferred $J_6$, $J_8$, the polar radius, MoI and the density distribution. 
We next explore the relation between the $J$ values, the planetary shape, and the depth of the winds. We demonstrate that accurate determinations of $J_6$ and $J_8$ can constrain the depth of the winds. 
We also show that more accurate measurements of Neptune's $J_2$ and $J_4$ can significantly reduce the possible parameter-space of solutions.  
We also present the rotation periods of Uranus and Neptune that are most consistent with the estimated polar radii. 
Finally, we investigate how an accurate measurement  of the MoI can constrain  the depth of the winds and rotation periods of Uranus and Neptune. 

Our main results can be summarized as follows:
\begin{enumerate}[leftmargin=1.5\parindent]
    \item The prediction of Uranus' and Neptune's $J_6$ and $J_8$ depend strongly on the dynamics. An accurate measurement of $J_6$ or $J_8$ (with a relative uncertainty of a few percents) is required to constrain the depth of the winds in Uranus and Neptune. 
    \item The density distribution in the deep interiors of Uranus and Neptune depends significantly on the rotation period and is strongly affected by dynamics.  
    For models assuming uniform rotation,  it is crucial to use wind-corrected gravitational coefficients: $J_{2n}^{stat} = J_{2n}^{meas} - J_{2n}^{dyn}$.
    \item More accurate measurements of $J_2$ and $J_4$ can further constrain the density distribution and narrow the range of predicted solutions in $J_6$, $J_8$ and MoI of Uranus and Neptune. 
    \item For both Uranus and Neptune  accurate determinations of the MoI could be used to distinguish between different  rotation periods and constrain the depth of the winds. For the former, a relative precision of 1\% is needed, whereas for the latter a relative precision of 0.1\%, is required.  
    \item We show that the generally used shapes of Uranus and Neptune do not agree with the broadly used rotation period $P_{\text{Voy}}$ and $P_{\text{HAS}}$. We, hence, reiterate the necessity of a robust and independent measurement of the rotation periods and shapes of Uranus and Neptune. 

\end{enumerate}
Future work could use the inferred pressure-density relations to interpret the presented empirical structure models in terms of composition. This would allow us to explore the possible compositions of Uranus and Neptune as well as to identify composition gradients and determine their dependency on the assumed rotation period and depth of the winds. We hope to address this in a future research.  

Finally, we emphasize the need for a dedicated space mission to Uranus and/or Neptune \citep[e.g.,][]{Arridge2014,Masters2014,Mousis2018,Hofstadter2019,Fletcher2020}.
We suggest that such a mission should be designed to measure the gravitational field of the planets, decreasing the uncertainty for $J_2$ and $J_4$, and determining the higher order $J_6$ and $J_8$ and the planetary shape, and if possible, the planetary rotation period, and MoI. 

\section*{Acknowledgements}
We thank N.~Movshovitz for many fruit-full discussions and technical support. We also acknowledge comments by the anonymous reviewer and support from the Swiss National Science Foundation (SNSF) under grant \texttt{\detokenize{200020_188460}}.

\section*{Data Availability Statement}
The data underlying this article will be shared on reasonable request to the corresponding author.

\bibliographystyle{mnras}
\bibliography{bib}

\appendix      

\section{Shape and gravity induced rotation period}
\renewcommand{\thefigure}{B1}
\begin{figure*}
    \centering
    \includegraphics[width = 0.5\textwidth]{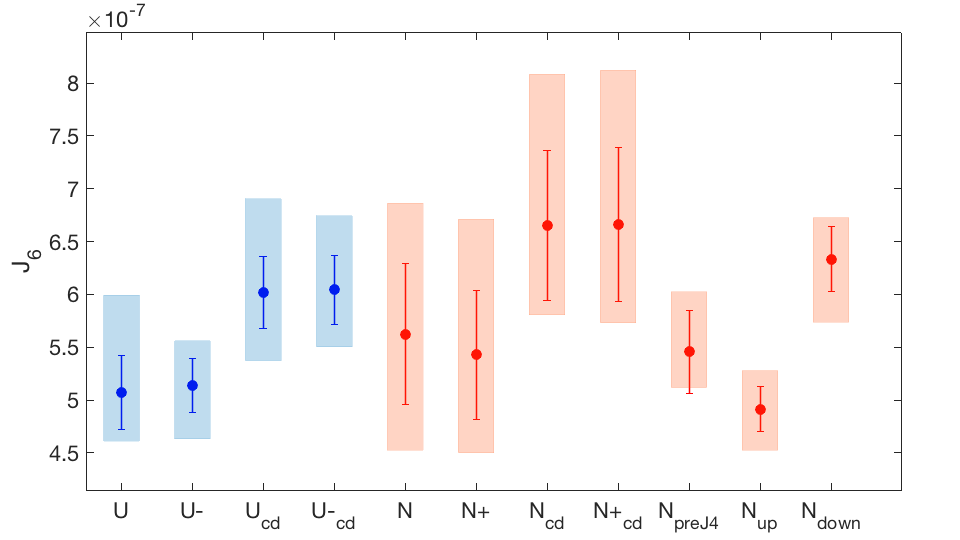}%
    \includegraphics[width = 0.5\textwidth]{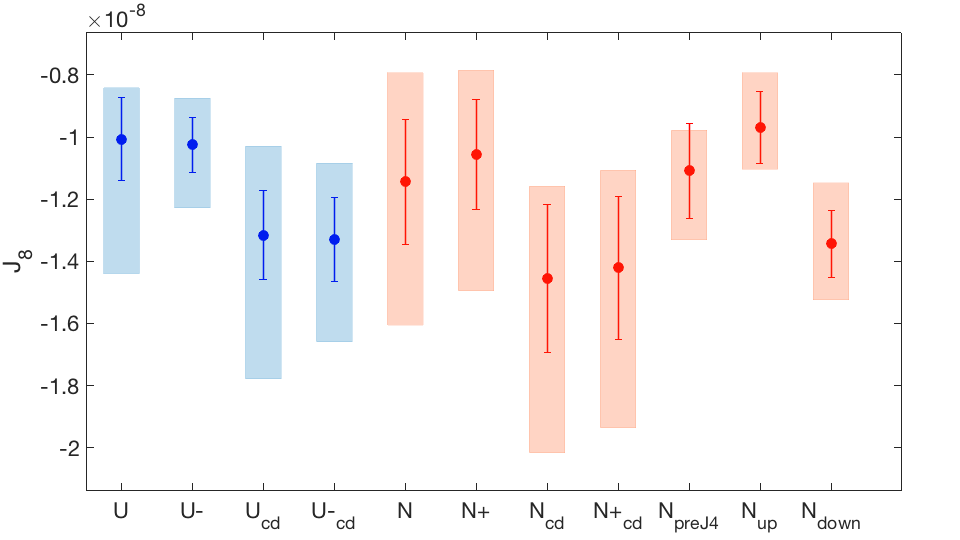}
    \includegraphics[width = 0.5\textwidth]{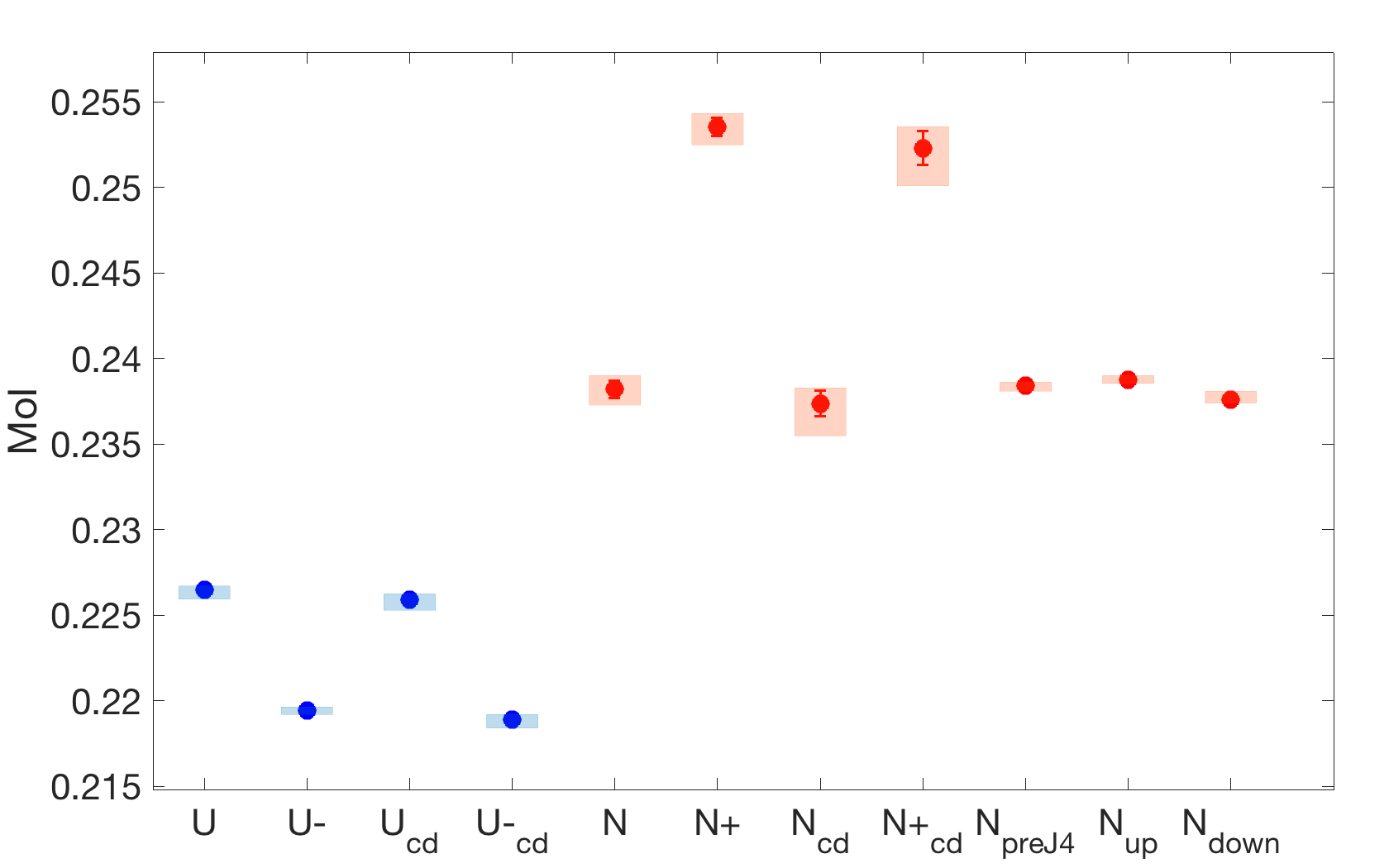}%
    \caption{Predictions of $J_6$ (left panel) and $J_8$-values (right panel) for different \textit{planetary models} of Uranus (blue colored) and Neptune (red colored). The \textit{planetary models} are collected along the x-axis with abbreviated names ("U": Uranus, "N": Neptune, "cd": corr dyn). The dots mark the mean values, bars the standard deviation, and boxes show the full solution range.}
    \label{fig:fancy_plots_ext}
\end{figure*}
Assuming uniform rotation and that the planet is in hydrostatic equilibrium, \cite{Helled2010_prot} showed that the flattening of Uranus and Neptune are inconsistent with the Voyager rotation periods ($P_{\text{Voy}}$). They expanded equation \ref{eq:total_potential} to second order and set equal the total potential on the equatorial radius $U(r,\phi=0)$ with the total potential on the polar radius $U(r,\phi=\pi/2)$. 
Then the expression can be solved for the rotation rate $\omega$. As $\omega=2\pi/\tau^{(2)}$, $\omega$ can be converted into the second-order rotation period $\tau_{}^{(2)}$ that is associated with the planetary shape and gravity field $J_2$ and $J_4$. \cite{Helled2010_prot} suggest that $\tau_{\text{u}}^{(2)} \approx$ 15 h 37 min 12 s and $\tau_{\text{n}}^{(2)} \approx$ 16 h 51 min 0 s for Uranus and Neptune, respectively.

Note that the rotation period of 16~h 51~min 0~s for Neptune as estimated in \cite{Helled2010_prot} is inconsistent with the shape induced rotation period estimation of 16~h 36~min 5~s as used here (see section \ref{subsec:constraining_power_of_shape}).
This is due to different assumed equatorial radii: in section \ref{subsec:constraining_power_of_shape} Neptune's equatorial radius has been adapted for each assumed rotation period (24,773.6~km for $P = 16$~h 36~min 5~s), while \cite{Helled2010_prot} use a constant value of $a=24,766$ km. The results, however, agree perfectly, when the radii are adapted to match. \\
We expand the formalism of \cite{Helled2010_prot} to fourth order, which is given by: 
\begin{align} \label{eq:rotation_period_J8}
    \tau^{(4)} &= 2\pi\sqrt{\frac{a^3}{2GM}}\left(\left(\frac{a}{b_0}\right)-1-J_2\left[\left(\frac{1}{2} \right) + \left(\frac{a}{b_0}\right)^3\right]\right.
 \nonumber \\
    &\left.-J_4\left[\left(-\frac{3}{8}\right)+\left(\frac{a}{b_0}\right)^5\right] -J_6\left[ \left(\frac{5}{16}\right) +\left(\frac{a}{b_0}\right)^7 \right]\right. \\
    &\left.- J_8 \left[ \left(-\frac{35}{128}\right)+ \left(\frac{a}{b_0}\right)^9\right]\right)^{-1/2}, \nonumber
\end{align}
and use $a$, $J_2$ and $J_4$ from Table \ref{tab:planet_properties} and mean values of $J_6$ and $J_8$ from Table \ref{tab:results} to infer the rotation period. We took for Uranus $GM = 5793951.3$ km$^3 \cdot$s$^{-2}$ \citep{Jacobson.2014} and $b_0 = 24,973$ km \citep{Archinal2018} and for Neptune $GM = 6835100.0$ km$^3 \cdot$s$^{-2}$ \citep{Brozovic2020} and $b_0 = 24,341$ km \citep{Archinal2018}. \\
For for $\tau_{}^{(2)}$ we get 15 h 38 min 59 s for Uranus and 16 h 50 min 08 s for Neptune. The small differences with respect to \cite{Helled2010_prot} arise mainly due to updated values in $GM$.
For $\tau_{}^{(4)}$ we get 15 h 39 min 0 s for Uranus and 16 h 50 min 10 s for Neptune, respectively. 
We confirm the results of \cite{Helled2010_prot} and show for both planets that $\tau^{(2)}$ and $\tau^{(4)}$ are nearly identical.
We therefore conclude that $J_6$ and $J_8$ do not refine $\tau^{(2)}$ for Uranus and Neptune.

\renewcommand{\thefigure}{C1}
\begin{figure*}
    \centering
    \includegraphics[width = 0.5\textwidth]{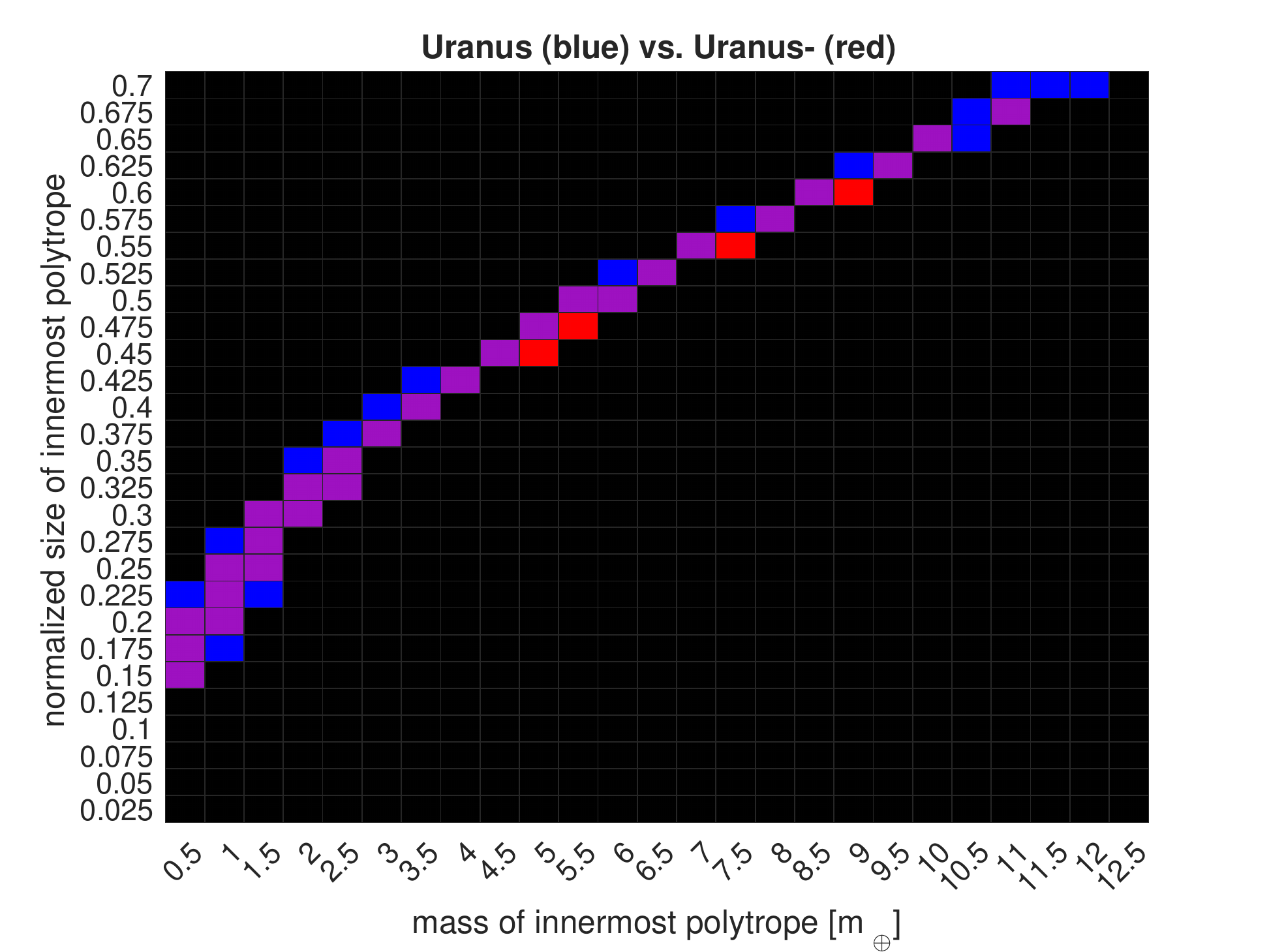}%
    \includegraphics[width = 0.5\textwidth]{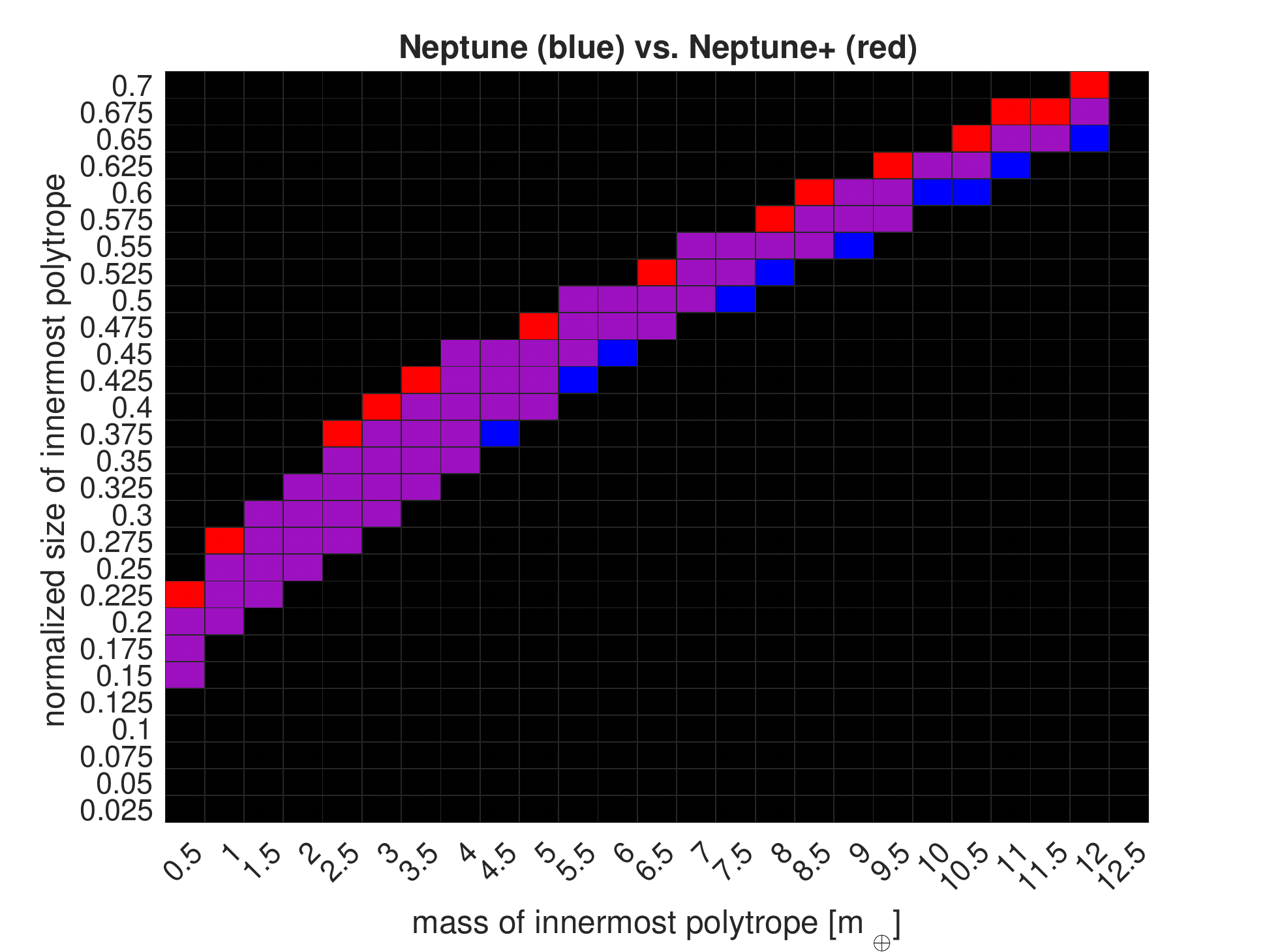}
    
    \includegraphics[width = 0.5\textwidth]{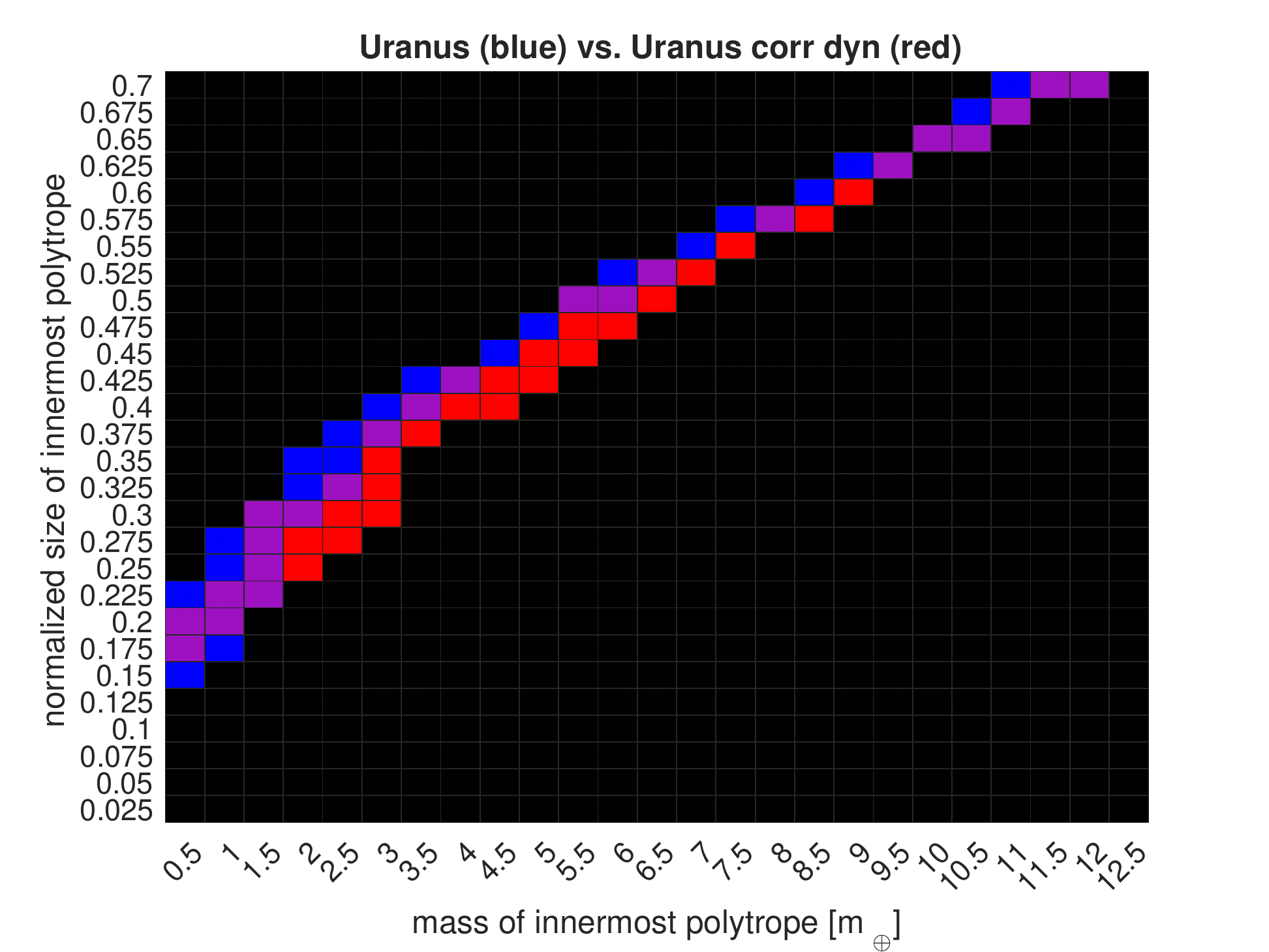}%
    \includegraphics[width = 0.5\textwidth]{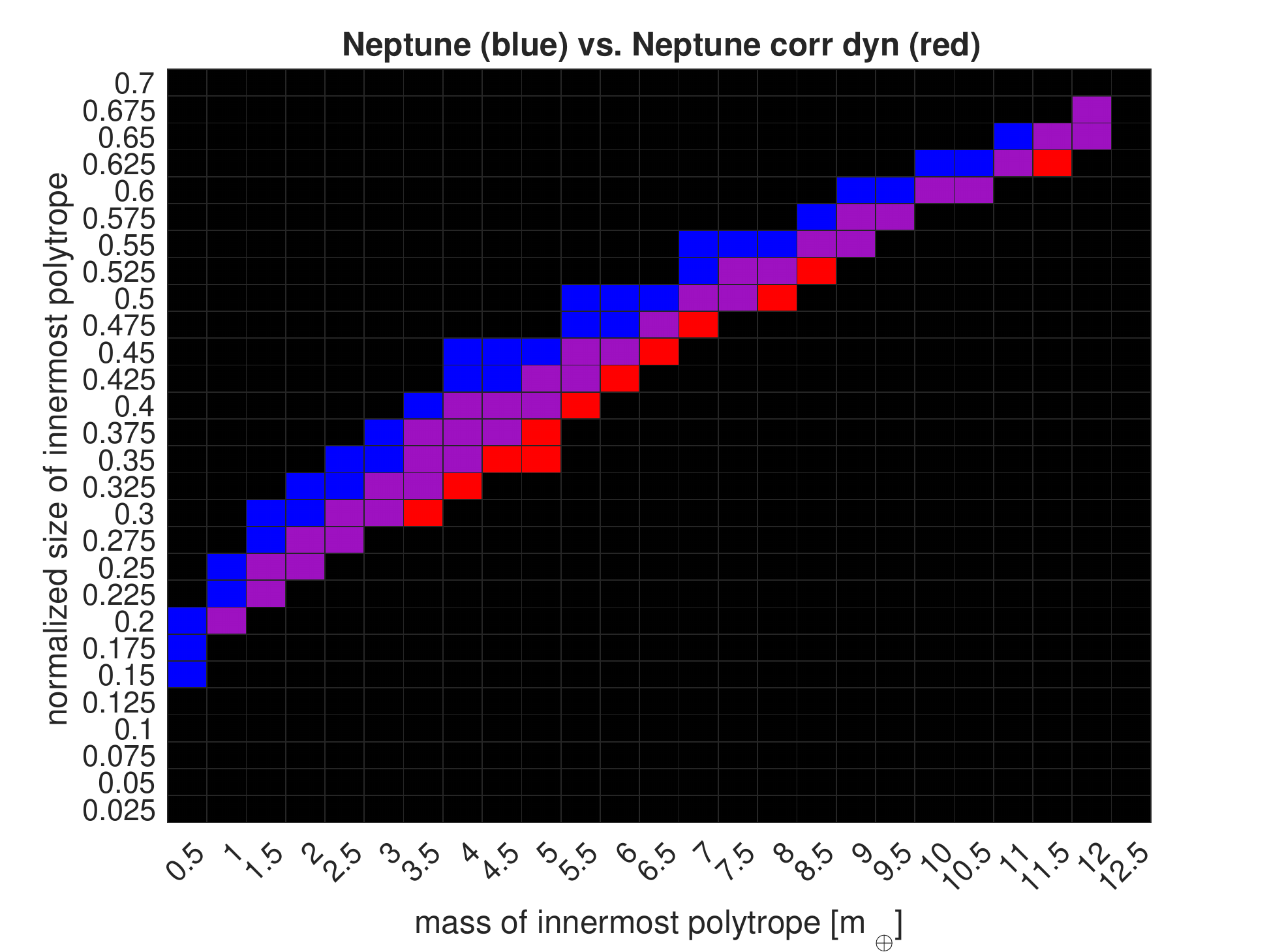}
    
    \caption{Solution spaces of \textit{planetary models} expressed in mass and size of the innermost polytrope. ($m_{\text{iP}}$, $r_{\text{iP}}$)-tuples for which a \textit{good solution} is found are colored. For tuples referring to black rectangles, no \textit{good solution} is found. Each panel shows the solution spaces of two \textit{planetary models} (indicated in the titles). 
    Depending in which \textit{planetary model} a \textit{good solution} is found for a specific tuple, the rectangle is colored in either blue or red (described in the titles). 
    A rectangle is colored in purple, if for the tuple a good solution is found in both \textit{planetary models}. }
    \label{fig:solution_space}
\end{figure*}

\section{Artificially improved $J_2$ and $J_4$ estimates} \label{sec:improvedJs}
In chapter \ref{subsec:better_known_J2_J4} we artificially improve the uncertainty of Neptune's $J_2$ by 85\% and $J_4$ by 75\% around the estimated values of \cite{Jacobson.2009}. We implicitly assume that the real values of $J_2$ and $J_4$ are close to the mean reported values \citep{Jacobson.2009} and, more specifically, within $J_{2,pre} = (3535.94 \pm 0.7)\cdot 10^{-6}$ and $J_{4,pre} = (-35.95 \pm 1.3)\cdot 10^{-6}$, respectively.
This, however, is not necessarily the case: the true value of $J_2$ and $J_4$ could be anywhere within the reported range, including near the boundaries of the uncertainty range. Such values are not included in the uncertainty range presented above. \\ 
Here, we investigate the solution space in $J_6$, $J_8$ and MoI of "Neptune $J_\text{pre}$" assuming different $J_{2,pre}$ and $J_{4,pre}$ value ranges. Self-evidently, all $J_{2,pre}$ and $J_{4,pre}$ value ranges are comprised in the uncertainties of $J_2$ and $J_4$, as estimated by \cite{Jacobson.2009}).
Concretely, we investigate two additional models. In the first model, the values of $J_{2,pre}$ and $J_{4,pre}$ are set to be as large as possible: $J_{2,pre} = (3539.74 \pm 0.7)\cdot 10^{-6}$ and $J_{4,pre} = (-34.35 \pm 1.3)\cdot 10^{-6}$. We call this planetary model "Neptune up". In the second model, the $J_2$ and $J_4$ value are set to be as low as possible: $J_{2,pre} = (3532.14\pm 0.7)\cdot 10^{-6}$ and $J_{4,pre} = (-37.55 \pm 1.3)\cdot 10^{-6}$. We call this planetary model "Neptune down". \\
Figure \ref{fig:fancy_plots_ext} shows the solution spaces in $J_6$ (top left panel), $J_8$ (top right panel) and MoI (bottom panel) for all \textit{planetary models}. The dots mark the corresponding mean, the error bars the standard deviation, and the colored boxes the full value range of each \textit{planetary model} for Uranus and Neptune. \\
We find that the predicted $J_6$, $J_8$ and MoI value ranges strongly depend on the chosen range of $J_{2,pre}$ and $J_{4,pre}$ values, incorporated in "Neptune $J_\text{pre}$", "Neptune up" and "Neptune down". \\
We conclude that in order to artificially improve the uncertainty in $J_2$ and $J_4$, not only one value range of more precise $J_{2,pre}$ and $J_{4,pre}$ can be considered. In fact, the whole existing uncertainty range of $J_2$ and $J_4$ has to be covered in order to not get biased towards one subset of $J_{2,pre}$ and $J_{4,pre}$.

\section{the solution space of the innermost polytrope} \label{sec:solution_space}
Here, we present the solution space of various \textit{planetary models} of Uranus and Neptune in terms of mass and size of the innermost polytrope. 
Figure \ref{fig:solution_space} shows the relation between the mass $m_{\text{iP}}$ and radius $r_{\text{iP}}$ of the innermost polytrope. If a \textit{good solution} is found for a certain ($m_{\text{iP}}$, $r_{\text{iP}}$)-tuple, the corresponding rectangle is colored.
For a tuple corresponding to a black rectangle, no solution is found. 
Each panel shows the solution spaces of two \textit{planetary models} (as indicated in the titles).
Depending in which \textit{planetary model} a \textit{good solution} is found, 
the corresponding rectangle is colored in either blue or red (described in the titles). The upper right panel for example draws \textit{good solutions} of "Neptune" in blue and \textit{good solutions} of "Neptune+" in red.
A rectangle is colored in purple, if for the tuple a \textit{good solution} is found in both \textit{planetary models}.

In the upper two panels of Figure \ref{fig:solution_space}, we observe that faster rotating planets tend to have smaller innermost polytropes (for a fixed $m_{\text{iP}}$) or have more massive innermost polytropes (for a fixed $r_{\text{iP}}$). This is in agreement with our findings in section \ref{sec:diff_rot_rate}. \\
The lower panels of Figure \ref{fig:solution_space} compare the solution spaces of \textit{planetary models} with wind corrected $J_4$ values ("Uranus corr dyn" and "Neptune corr dyn") to "Uranus" and "Neptune", respectively. We observe that the solution spaces of "Uranus corr dyn" and "Neptune corr dyn" include more massive innermost polytropes (for a fixed $r_{\text{iP}}$) or smaller innermost polytropes (for a fixed $m_{\text{iP}}$). This is in agreement with section \ref{subsec: effect of dynamics}.

Note that, compared to Neptune, the solution spaces of Uranus models are in general more constraint. This is due to Uranus' more precise gravity data and is in agreement with section \ref{sec:results} and \cite{NETTELMANN2013}.

\section{(known) dependencies of the gravity field} \label{section:polar_radius} \label{sec:graveyard}
In this section we present the  relation between the planetary  shape and $J_2$, and the correlations between $J_4$, $J_6$ and $J_8$. 
The planetary flattening is related to its $J_2$ value \citep[e.g.,][]{Mecheri2004, Helled2011}. \\ Figure \ref{fig:J2_polar_radius} shows the relation between Neptune's polar radius and $J_2$ value. Neptune has been chosen as a representative for all \textit{planetary models}. The dashed line marks the best fitting curve.
We note that an accurate measurement of the polar radius can further constrain Neptune's $J_2$ and vise-versa. However, for Neptune, a relative accuracy in either $b_0$ of 10$^{-5}$ or in $J_2$ of 10$^{-3}$ is needed to further constrain the other. \\
\renewcommand{\thefigure}{D1}
\begin{figure}
    \centering
    \includegraphics[width = 0.5\textwidth]{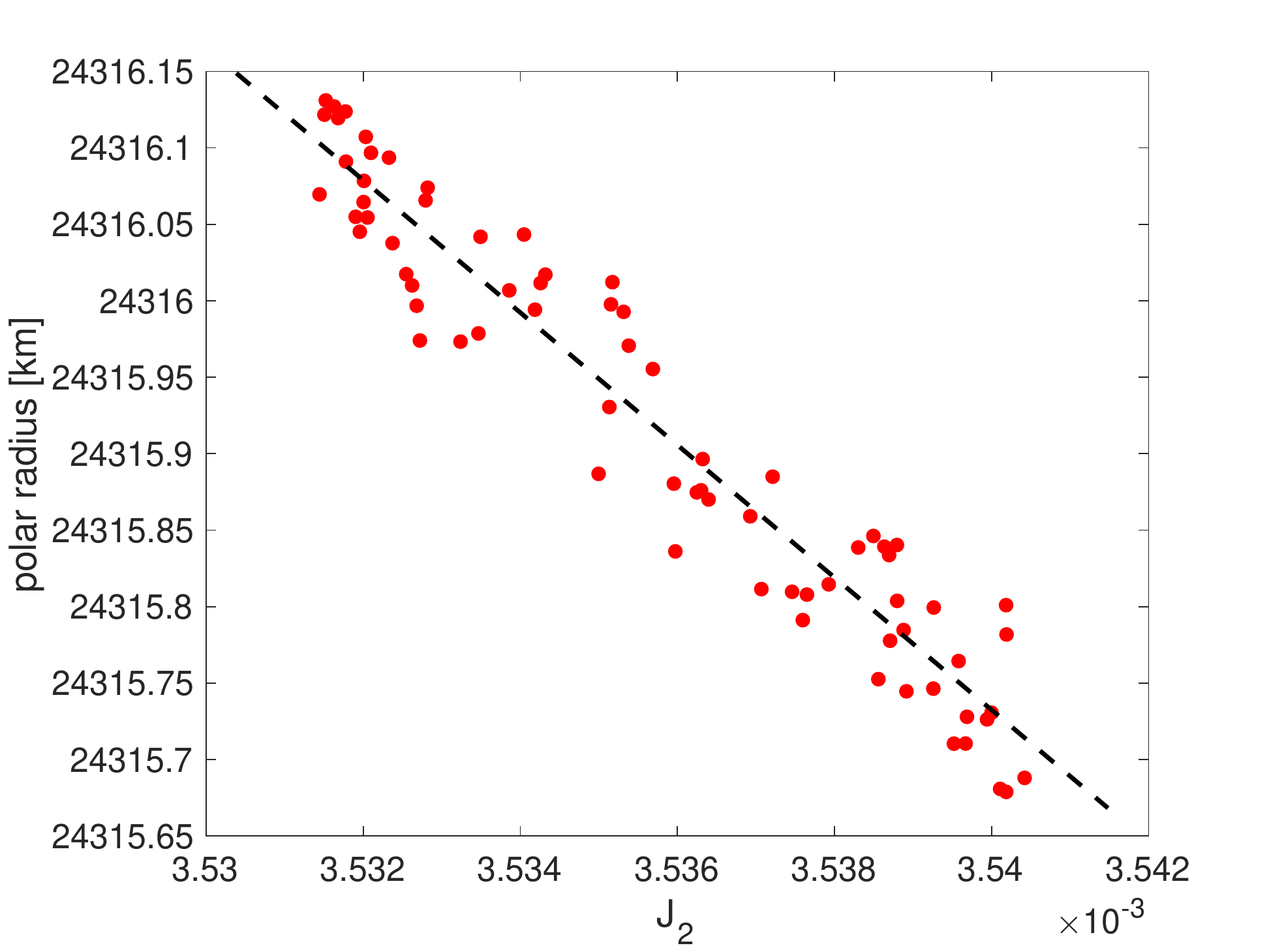}%
    \caption{Neptune's polar radius depending on its $J_2$ value. The dotted line marks the best-fitting curve.}
    \label{fig:J2_polar_radius}
\end{figure}
Figure \ref{fig:J4_vs_J6_vs_J8} shows the relation between $J_4$ (x-axis), $J_6$ (y-axis) and $J_8$ (color-coded) for "Neptune+", as a representative for all \textit{planetary models}. We observe a clear relation between $J_4$ and $J_6$, and a clear trend in color. This suggests that an accurate estimate of any of $J_4$, $J_6$, or $J_8$ can constrain the remaining ones.

\renewcommand{\thefigure}{D2}
\begin{figure}
    \centering
    \includegraphics[width = 0.5\textwidth]{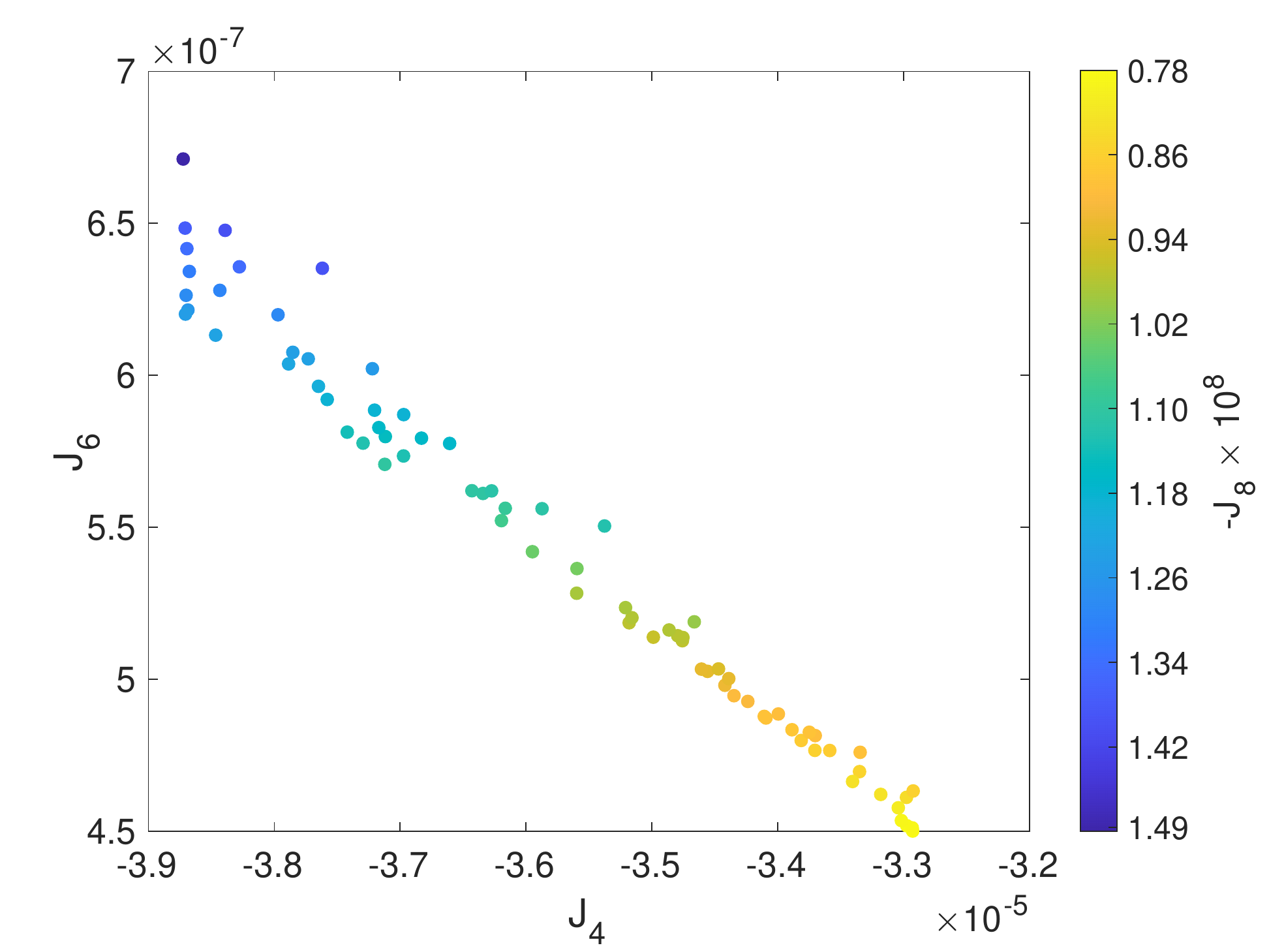}%
    \caption{Relation between $J_4$, $J_6$ and $J_8$ of "Neptune+", which is a representative of all \textit{planetary models}. The color is indicating the $J_8$ value of each solution.}
    \label{fig:J4_vs_J6_vs_J8}
\end{figure}

\section{polytropes} \label{subsec:polytropes}

A polytrope relates the pressure $P$ to the density $\rho$ (see equation \ref{eq:polytrope}). $n$ is the polytropic index and $K$ the polytropic constant. 
$n$ and $K$ define the relation between $P$ and $\rho$, which in turn defines the resulting density profile. \\
In this chapter we quantify relations between the polytropic index $n$ and the planet's density profile.

We generate internal structure models consisting of three piece-wise arranged polytropes. 
We assign $n_1$ and $K_1$ to the outermost polytrope. It defines the region between the surface and $r_{trans}$. $n_2$ and $K_2$ is assigned to the second polytrope that defines the intermediate region between $r_{trans}$ and $r_{core}$. Finally, $n_3$ and $K_3$ are assigned to the innermost polytrope that represents the deep interior (reaching from $r_{core}$ to the center of the planet). 

In general, the polytropic index $n$ is related to the "curvature" of the density profile. In other words, a larger $n$-value results in a steeper decrease in density. $K$, on the other hand, is related to the offset of the density profile: a larger $K$ lowers the overall density of the corresponding polytropic region. \\
Changing $K_i$ and/or $n_i$ does not only affect the density distribution in the corresponding polytropic region, but also in the whole planet. 
This is due to the calculation technique of our ToF-implementation: during the calculation of the hydrostatic equilibrium, the total planetary mass and radius are normalized. Only afterwards, the resulting density profile is up-scaled to match the total planetary mass. 
Changes in a polytrope change the density distribution in the corresponding polytropic region that may alter the up-scaling factor. A changed up-scaling factor finally scales differently the whole planetary density profile. \\
Figure \ref{fig:neptune_3_polytrope} shows how a change in $n_2$ affects the entire planetary density profile. The colors of the curves represent the $n_2$ value (see legend). We fix $K_1 = 100,000, n_1=0.6, K_2=50,000, K_3=45,000, n_3 = 0.9, r_{core} = 0.2, r_{trans} = 0.8$. We apply the polytropes on a Uranus-like planet.
Uranus-like means, this planet has the same mass, radius,  and rotation period but a different gravitational field than Uranus. \\
We observe that changes in $n_2$ do not only affect the region defined by the second polytrope, but the entire density profile. This includes the density discontinuities at $r_{core}$ and $r_{trans}$. The closer two neighboring polytropic indices, the smaller the density discontinuity between the two polytropic regions.

Although changes in each $K_{1,2,3}$ and/or $n_{1,2,3}$ change the density profile, not all changes are equal. First, changes in $n_{1,2,3}$ alters the density profile more substantially than changes in $K_{1,2,3}$. Second, different $n_2$ alter the density profile the most, followed by different $n_3$ and $n_1$. This is due to different enclosed masses in each polytropic region: while the outermost polytrope (associated with the planet's envelope region) encloses only little mass, the second polytrope generally incorporates most planetary mass. The more mass is affected by changing $n_i$, the larger the effect on the whole planetary density profile and its density discontinuities. Hence, changes in $n_1$ mainly affects the density in the envelope and the discontinuity at the transition radius $r_{trans}$. On the other hand, changing $n_2$ has mayor effects on both density discontinuities at $r_{trans}$ and $r_{core}$ and alters the whole density profile significantly (see Figure \ref{fig:neptune_3_polytrope}). 

We find that for $n_{i+1} \gsim n_{i}$ the density profile is monotonically decreasing towards the planet's surface (e.g., has no negative density heights at $r_{core}$ or $r_{trans}$). 
The zoom plot in Figure \ref{fig:neptune_3_polytrope} demonstrates that at $r_{core}$. For $n_2=0.92>n_3=0.9$, a negative density height at $r_{core}$ is observed (thicker line). 
No "strict" inequality sign is used in $n_{i+1} \gsim n_{i}$, however, as second order effects as $K_i$ still can avoid potential negative density jumps. 

\renewcommand{\thefigure}{E1}
\begin{figure}
    \centering
    \includegraphics[width = 0.5\textwidth]{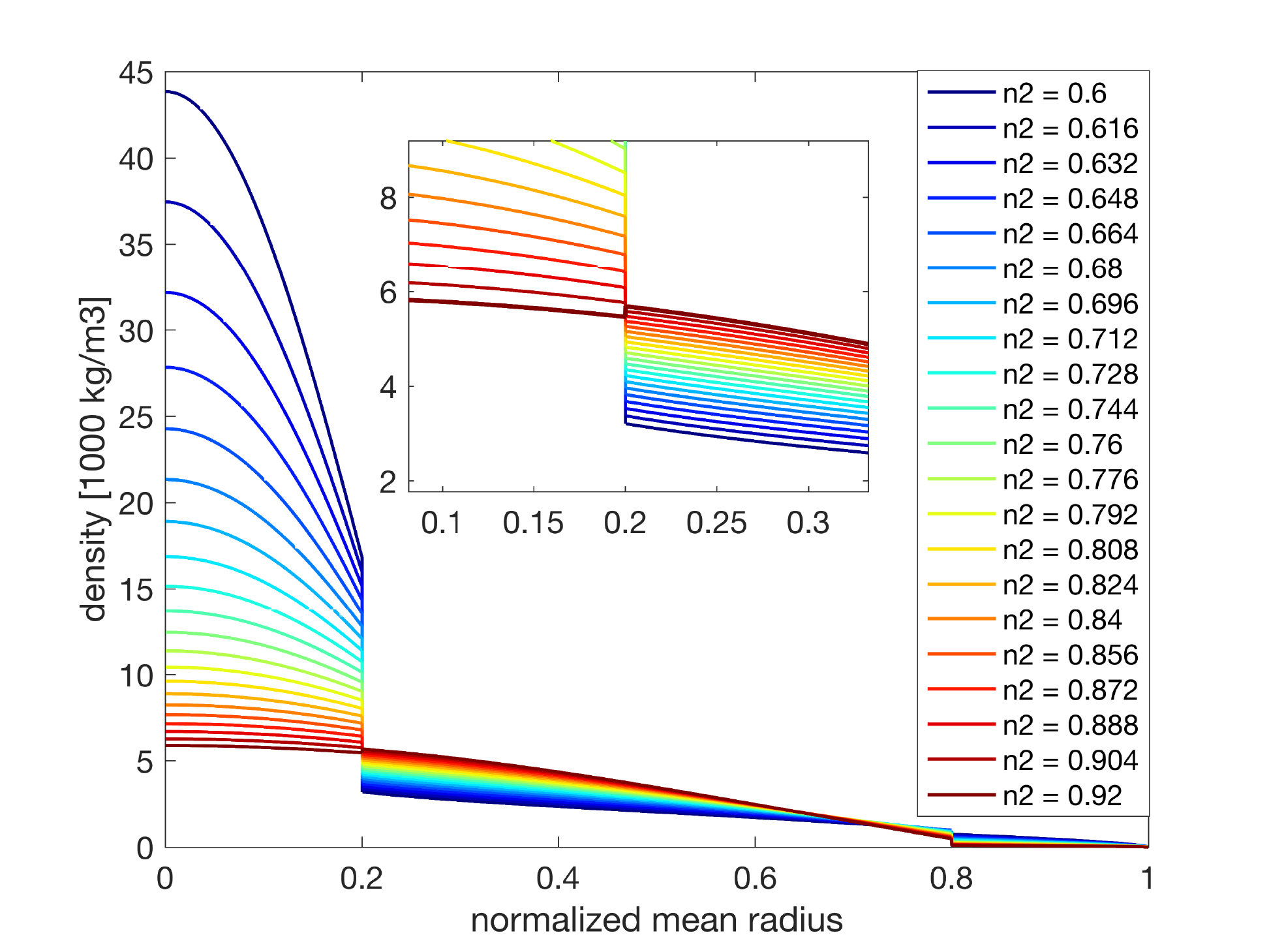}%
    \caption{Density profiles of a Uranus-like planet depending on only $n_2$. Each curve is colored according to its $n_2$ value. }
    \label{fig:neptune_3_polytrope}
\end{figure}

The upper panel in Figure \ref{fig:polytropic_indices}, shows the relation between $n_1$ (x-axis) and $n_2$ (y-axis). Colored dots show solutions of the \textit{planetary model} "Neptune", whereas black crosses mark solutions of "Uranus". The colors correspond to the size of the density discontinuity at $r_{trans}$. The dashed line marks $n_1 = n_2$. It is always true that $n_2>n_1$. This is expected, as we request our density profile to be monotonically decreasing. \\
The lower panel in Figure \ref{fig:polytropic_indices} shows the relation between $n_2$ (x-axis) and $n_3$ (y-axis). Again, colored dots and black crosses mark solutions of "Neptune" and "Uranus", respectively. The color corresponds to the size of Neptune's density discontinuity at $r_{core}$. The dashed line marks $n_2 = n_3$. \\
We observe that $n_{3} \approx n_{2}$. But we expect $n_3 \gsim n_2$ as we require the density profile to monotonically decrease with larger radius.
This behavior can be explained as follows. For all \textit{planetary models} we set an upper limit for the central density of $\rho_{central} \gsim 18,000$ kg$\cdot$m$^{-3}$. This in turn prevents that $n_{3} \gg n_{2}$, as this could induce large density discontinuities at $r_{core}$ and high central densities (see Figure \ref{fig:neptune_3_polytrope}). As a consequence, $n_2$ and $n_3$ tend to be rather similar. 
In cases where $n_2>n_3$, second order effects as e.g., large $K_2$ values prevents a negative density jump. 
Although similar, the solution spaces of "Neptune" in Figure \ref{fig:polytropic_indices} allow for a broader range in comparison to the solution spaces of "Uranus. This may again be a consequence of the larger uncertainties in the measured gravity field. 

Different polytropic indices correspond to different matter properties. If, therefore, with more accurate gravity data, Neptune's solutions no longer coincide with Uranus' solution space in either $n_1$, $n_2$ or $n_3$, one can conclude that Uranus and Neptune consist of different composition properties. This in turn would be another indication for the dichotomy of the two "Ice Giants".

\renewcommand{\thefigure}{E2}
\begin{figure}
    \centering
    \includegraphics[width = 0.5\textwidth]{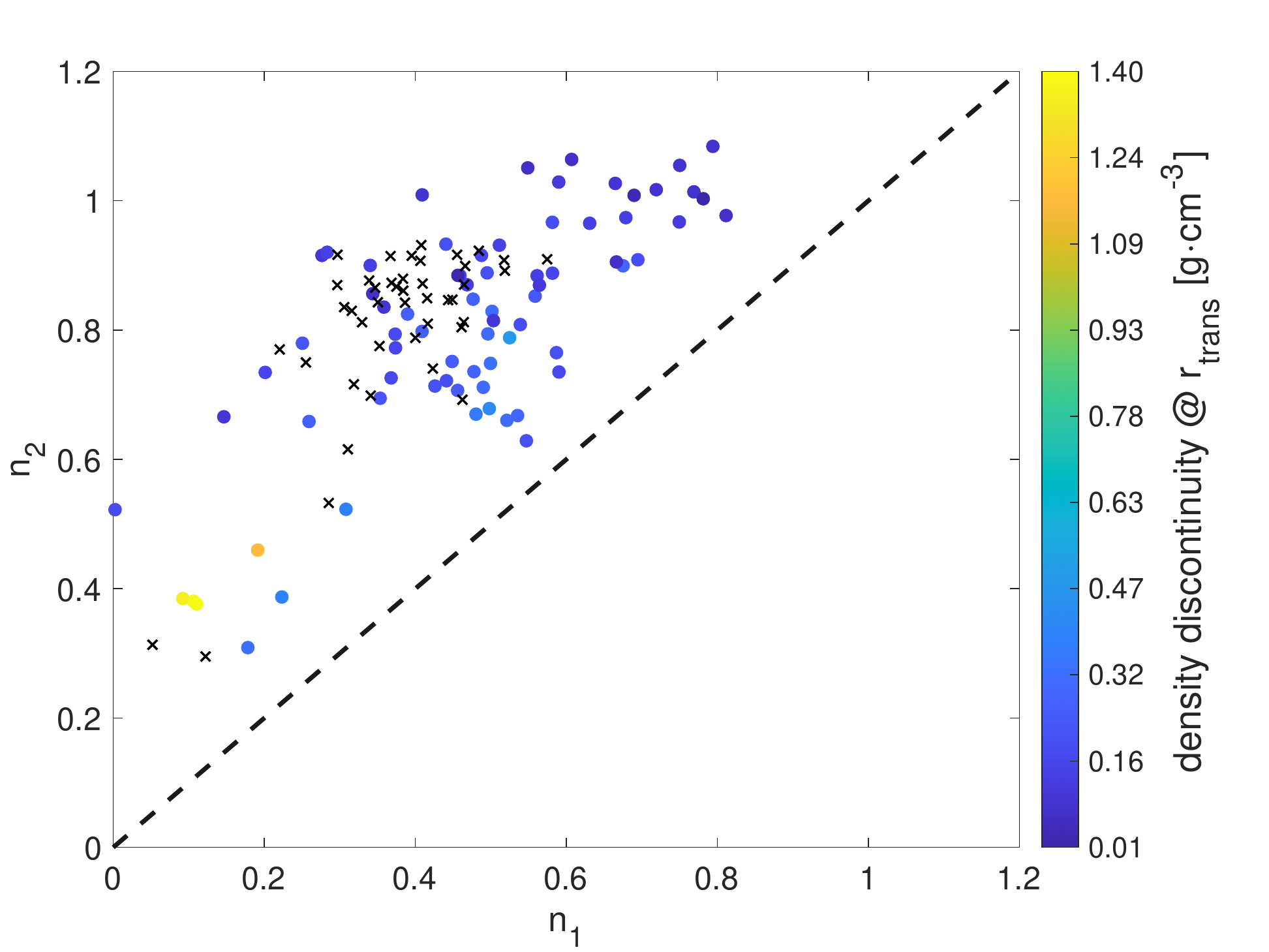}%
    
    \includegraphics[width = 0.5\textwidth]{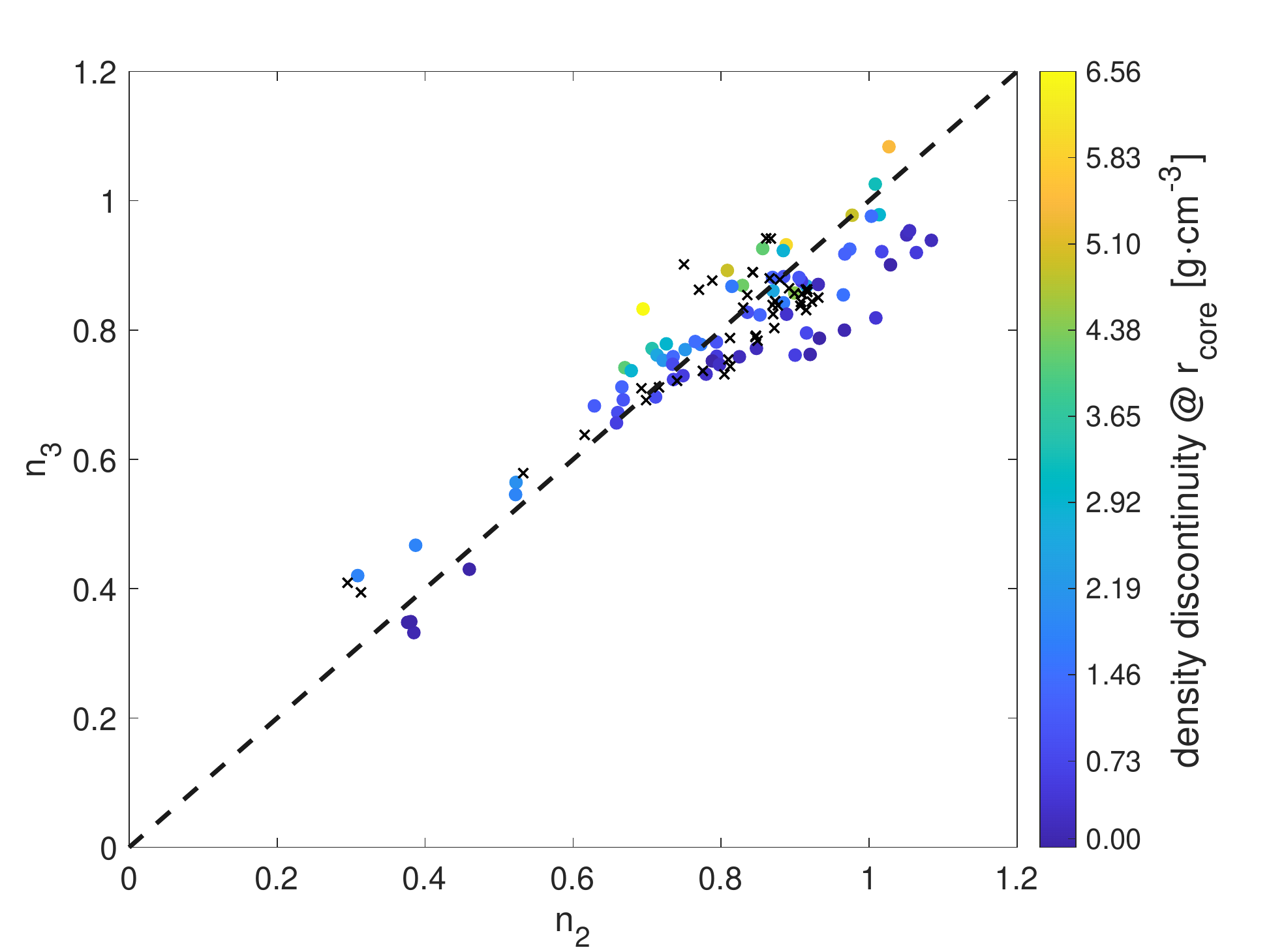}
    \caption{
    \textbf{Upper panel:} The relation between $n_2$ and $n_3$ for Neptune (colored points) and Uranus (black crosses). For Neptune, the color is representing the solution's density discontinuity height at the transition radius. The black dashed line shows $n_2 = n_3$ and helps to guide the eye. 
    \textbf{Lower panel:} The relation between $n_1$ and $n_2$ for Neptune (colored points) and Uranus (black crosses). In Neptune's case, the color is representing the solution's density discontinuity height at the core radius. The black dashed line represents $n_1 = n_2$ and helps to guide the eye. 
    }
    \label{fig:polytropic_indices}
\end{figure}

\end{document}